\documentclass[12pt]{spieman} 
\usepackage{amsmath,bm,amsfonts,amssymb}
\usepackage{graphicx}
\usepackage{setspace}
\usepackage{tocloft}
\usepackage{multirow}
\usepackage{upgreek}
\usepackage{lineno}
\usepackage{xcolor}
\title{Exoplanet Detection in Starshade Images}

\author[a,*]{Mengya (Mia) Hu}
\author[a]{Anthony Harness}
\author[b]{He Sun}
\author[c]{N. Jeremy Kasdin}
\affil[a]{ Department of Mechanical and Aerospace Engineering,
 Princeton University,
 Princeton, NJ, 08544, USA}
\affil[b]{Department of Computing and Mathematical Sciences,
 California Institute of Technology,
 Pasadena, CA, 91125, USA}
\affil[c]{College of Arts and Sciences,
University of San Francisco,
San Francisco, CA, 94117, USA}

\cftpagenumbersoff{figure}
\cftpagenumbersoff{table} 
\begin{document} 
\maketitle

\begin{abstract}
A starshade suppresses starlight by a factor of $10^{11}$ in the image plane of a telescope, which is crucial for directly imaging  Earth-like exoplanets.  The state of the art in high contrast post-processing and signal detection methods were developed specifically for images taken with an internal coronagraph system and focus on the removal of quasi-static speckles. These methods are less useful for starshade images where such speckles are not present. This paper is dedicated to investigating signal processing methods tailored to work efficiently on starshade images.  We describe a signal detection method, the generalized likelihood ratio test (GLRT), for starshade missions  and look into three important problems. First,  even with the light suppression provided by the starshade, rocky exoplanets are still difficult to detect in reflected light due to their absolute faintness. GLRT can successfully flag these dim planets. Moreover, GLRT provides estimates of the planets' positions and intensities and the theoretical false alarm rate of the detection. Second, small starshade shape errors, such as a truncated petal tip, can cause artifacts that are hard to distinguish from real planet signals; the detection method can help distinguish planet signals from such artifacts. The third direct imaging problem is that exozodiacal dust degrades detection performance. We develop an iterative generalized likelihood ratio test to mitigate the effect of dust on the image. In addition, we provide guidance on how to choose the number of photon counting images to combine into one co-added image before doing detection,  which will help utilize the observation time efficiently. All the methods are demonstrated on realistic simulated images. 

\end{abstract}

\keywords{Starshade, Image processing, Generalized likelihood ratio test, Exoplanet detection, High contrast imaging }

{\noindent \footnotesize\textbf{*}Mengya (Mia) Hu,  \linkable{mengyah@princeton.edu} }

\begin{spacing}{2}   

\section{Introduction}
\label{sect:intro} 
A sun-like star is much brighter (typically 100 million to 10 billion times) than an Earth-like planet in its habitable zone \cite{Traub2010}. Moreover, at a distance of 10 pc, the star and planets in its habitable zone are separated by around 0.1 arcseconds. Thus, it is difficult to separate the planet light from that of the star in the image. There are two main solutions to the challenge of imaging objects in close proximity to much brighter ones. First, one can use a coronagraph\cite{lyot}, which is a device inside the telescope to block the starlight from reaching the image plane. Second, one can use a starshade\cite{spitzer, Cash_2006}, which is a large screen flying on a separate spacecraft positioned between the telescope and the star being observed to suppress the starlight before it enters the telescope.  In many ways, coronagraphs and starshades are complementary. Coronagraphs are efficient  for high-contrast surveys, because it is easy to point the instrument to different targets. However, they  result in a lower optical throughput relative to a starshade, have the difficulty for designing a coronagraph due to `exotic pupils',  and are very sensitive to wavefront perturbations. Even small aberrations introduce bright speckles, which influence the instrument's ability for exoplanet observations\cite{wavefront}. Thus, it imposes many challenging requirements on the telescope and instruments to design coronagraphs with $\sim 10^{-10}$ starlight suppression; two mission concepts under study are Habitable Exoplanet Observatory (HabEx)\cite{HabEx} and Large UV/Optical/Infrared Surveyor (LUVOIR) \cite{LUVOIR}. In comparison, starshades are good at deep imaging and spectroscopic characterization. They are not sensitive to wavefront errors and can be designed to operate over a large bandpass. The total throughput is high since the starshade does not require any internal masking of the optical beam, which makes a starshade an excellent option for deep spectroscopy, especially at small inner working angles (IWA). However, one disadvantage of a starshade is the time it takes to slew the starshade to realign it with different target stars. The starshade's ability to efficiently suppress the on-axis starlight while maintaining high throughput makes it an excellent tool for exploring the habitability of exoplanets. A recently studied potential mission is the Starshade Rendezvous Mission: a starshade that will work with the Nancy Grace Roman Space Telescope (previously called WFIRST)\cite{sa2019}. In the mission, the starshade is launched separately and rendezvous with the telescope in orbit. Starshades are also baselined for the HabEx mission concept\cite{HabEx}. 

 Starshades are a new technology, still in development. Coronagraphs, however, have been used on ground-based telescope for decades; even the Hubble Space Telescope has a rudimentary coronagraph. Available research on high-contrast imaging is mostly about image processing and signal detection for coronagraph observations, which focus on alleviating the influence of quasi-static speckles. However, quasi-static speckles are not present in starshade images, so the emphasis is on the starshade's error sources. The dominant sources of noise in starshade images are sunlight scattering off the starshade edges\cite{Martin2013} and unsuppressed starlight caused by errors in the starshade shape\cite{Shaklan2010}. The scattered sunlight is confined to two extended lobes perpendicular to the direction of the sun and is constant during observations\cite{sunlight_lobes}. Its stability means it could potentially be calibrated out, just adding photon noise to a small region around the starshade. Manufacturing and deployment errors and thermal deformations can distort the starshade shape and will produce bright spots in images that are hard to distinguish from a real planet signal. One example shape error we examine in this study is the truncation of the tips of starshade petals. Additional sources of noise in starshade images come from misalignment of the starshade and telescope, detector noise, and exozodiacal dust\cite{dust}. 

 Due to the difference of the noise properties in coronagraph images and starshade images, previous work on coronagraphs loses its utility on starshade images, which motivates the investigation of new techniques for starshade images. In this paper, we focus on the impact caused by errors in the starshade shape and exozodiacal dust and present an automatic detection algorithm, the generalized likelihood ratio test (GLRT), to provide robust detection on low-signal images in the presence of shape errors. We have described the GLRT model and its preliminary results for simulated images with starshade shape errors, dynamics and detector noise in Ref.~\citenum{glrt}. We will review this detection method and introduce an iterative process to detect a planet in the presence of significant exozodiacal dust.  This work focuses on signal detection without the need for post-processing (e.g., PSF subtraction). Post-processing may improve the detection performance but could also  complicate the data analysis process and risks introducing artifacts into or removing part of the planetary signal. We believe demonstrating our signal detection method on unprocessed images strengthens the argument for the efficiency of our method.   Detailed investigation on post-processing is beyond the scope of the paper.

In Sec.~\ref{sec:simulation}, we describe the image simulation process used to generate the test set for our planet detection methods.  Sec.~\ref{sec:glrt} presents the GLRT detection method and represents the bulk of this work. An iterative approach of GLRT is presented in Sec.~\ref{subsec:dust} to tackle the problem of exozodiacal dust. We end by summarizing our work. 

\section{Image generation} \label{sec:simulation}
\begin{figure}[ht!]
\begin{center}
\begin{tabular}{c}
\includegraphics[height=6.3cm]{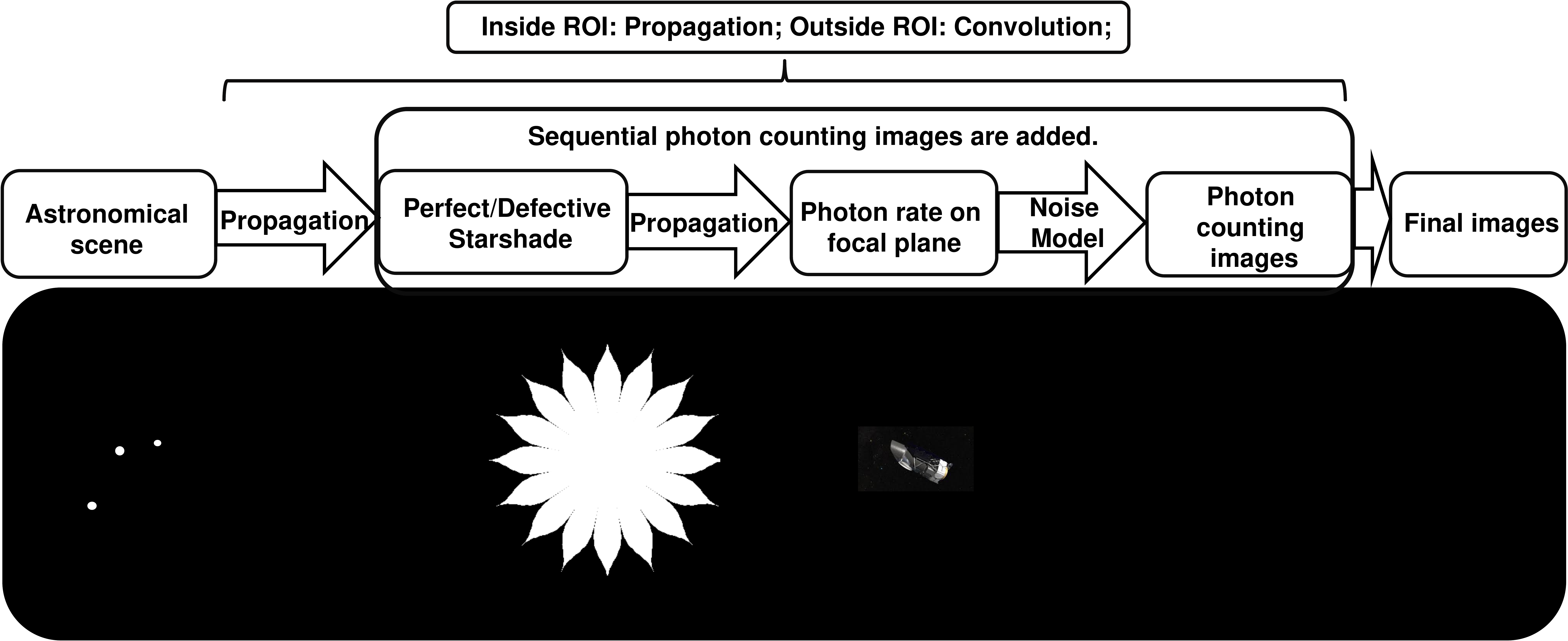}
\end{tabular}
\end{center}
\caption{Diagram of the image simulation process with starshade system illustration   (not drawn to scale). The illustration, i.e., light sources, a starshade and a telescope, on the bottom is aligned with the description above. ROI is the area defined at the simulation input, i.e., the astronomical scene, beyond which the incoming light is considered a plane wave and is not diffracted by the starshade.}    \label{fig:generation}
\end{figure}

We briefly summarize the image generation process outlined in Fig.~\ref{fig:generation}. A more detailed treatment can be found in Ref.~\citenum{simulation}. The input for the image generation process is a model of the solar system viewed face-on from 10 pc away developed as part  of the Haystacks Project \cite{haystack}. This model contains multi-wavelength image slices, covering the range from 0.3 $\upmu$m to 2.5 $\upmu$m. Each image centers on the star, extending to 50 AU from it. The star and planets are represented as single-pixel sources  and the model also includes interplanetary dust and astrophysical background sources. The pixel values in the image slices are spectral flux densities.

\begin{figure}
\begin{center}
\begin{tabular}{c}
\includegraphics[width=0.5\textwidth]{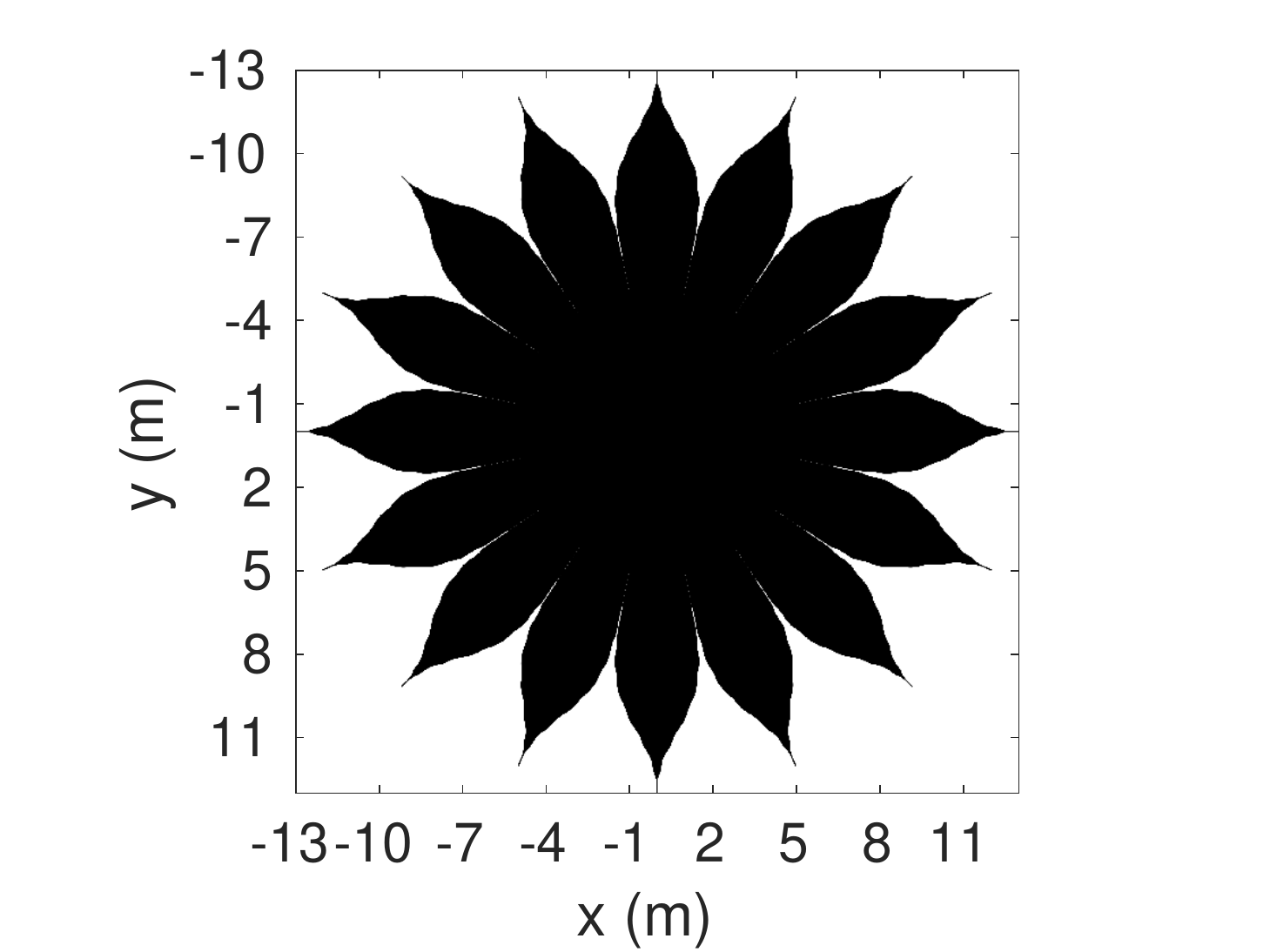}
\end{tabular}
\end{center}
\caption{The starshade we use in this paper, which is 13 meters in radius and has 16 petals  and designed for the Starshade Rendezvous Mission\cite{sa2019}. \label{fig:ss}}
\end{figure}

The optical model to calculate the starshade diffraction uses Fresnel diffraction theory\cite{fresnel} to propagate light past the starshade. However, calculating the propagation of each pixel separately in Haystacks is computationally expensive. Starshades have a noticeable influence on light propagation only for a small area in close proximity to the starshade, which we call the region of influence (ROI).  The ROI is defined at the input plane of the simulation and is the angular separation of a source, beyond which the incoming light is considered a plane wave and is not diffracted by the starshade. The result for an off-axis light source outside the ROI is close to the result for the same light source as if there were no starshade. Thus, we only use Fresnel propagation inside the ROI and simply convolve point sources outside the ROI with the telescope's point spread function (PSF). The starshade we use in this paper is 13 meters in radius and has 16 petals, shown in Fig.~\ref{fig:ss}.

The image simulation incorporates the main factors that influence the image of a realistic system:  a binary matrix represents the starshade outline; defects are added to the shape by adjusting this matrix; the alignment between the starshade and telescope is accounted for by adding time-dependent formation flying dynamics to the diffraction calculation; and a detector model for the Roman Telescope is used, which includes clock-induced charge, dark current, degradation during lifetime, polarization losses, and read noise \cite{detector}. 

Planets are extremely faint, so their signals can be weak compared to the readout noise. To tackle this problem, the Roman Telescope uses an electron-multiplication charge-coupled device (EMCCD)\cite{detector}, which amplifies the signal in an electron-multiplication register. This process reduces the effective readout noise to less than one electron \cite{PCmodel}. EMCCDs introduce an additional noise source: the multiplicative noise associated with the amplification process, which can be eliminated by using the detector in photon counting (PC) mode. PC mode applies a chosen threshold to the number of electrons generated in each pixel and yields a value of one if the number of electrons is larger than the threshold, zero otherwise. It is a binary process that does not measure the exact number of photons, but rather the presence of photons.  As PC mode can not distinguish the event of one photon from the event of more than one photon,  the exposure time is short enough so that the expected number of photons in any pixel  of the detector is less than one. In this way, photons aren't wasted. In this work, we choose the integration time for each PC image to be 1 second, so that the maximum photon rate on the detector (looking at an Earth-like planet from 10 pc) is around 0.1 photon per second per pixel. 

\begin{figure}
\begin{center}
\begin{tabular}{c}
\includegraphics[width=0.97 \textwidth]{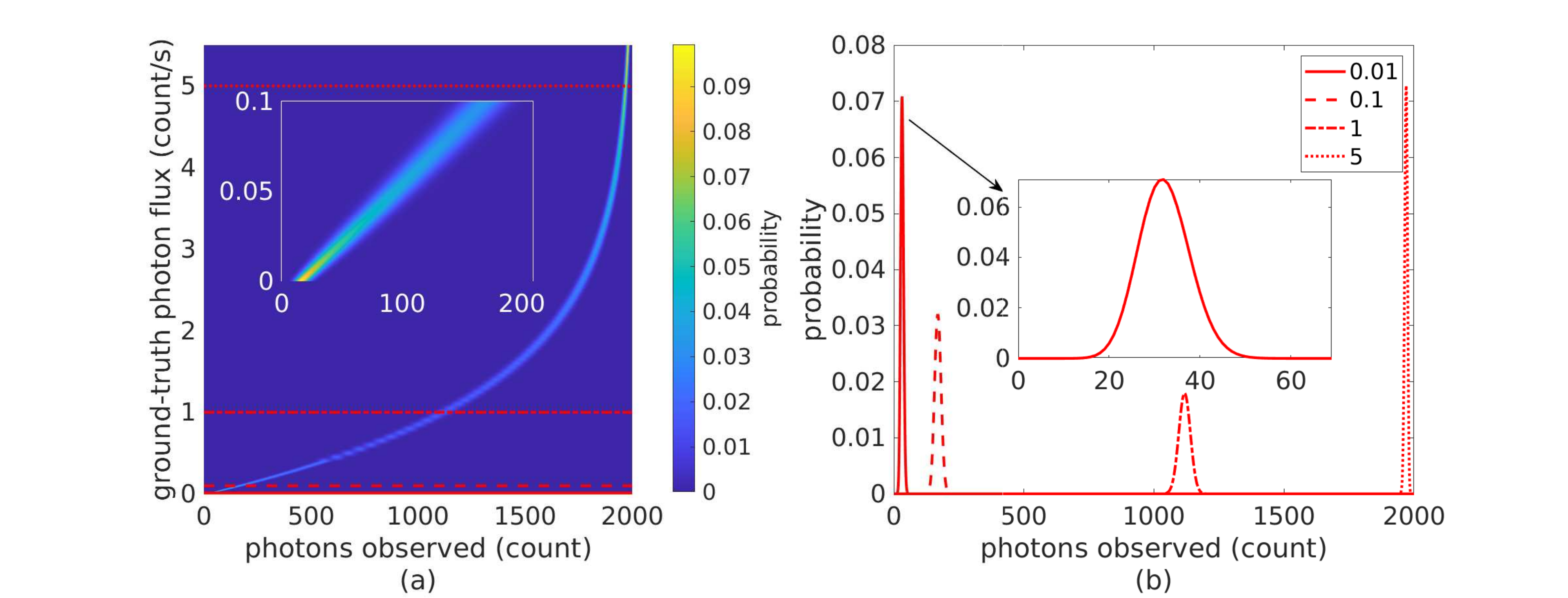}
\end{tabular}
\end{center}
\caption{   Probability density functions of the photon counts for different ground-truth photon flux. (a). Each row is the probability density function (PDF) of photon counts observed in one pixel in a co-added image from 2000 PC images corresponding to the photon flux in that pixel. (b). Example PDF's of the photon counts for four different ground-truth photon flux, which correspond to the ones in (a). The PDF of photon counts observed in one pixel in a co-added image from 2000 PC images corresponding to the photon flux 0.01  count/s, 0.1 count/s, 1  count/s and 5 count/s in the pixel. \label{fig:pdf}}
\end{figure}

Typically, the `binary' PC images are not used directly, but rather are added together to create a final image, which we will call a co-added image. In this way, the co-added images have large enough dynamic range so the different intensity of the sources can be well-reflected. Most of the images shown in this paper are co-added final images, unless otherwise stated. The number of PC images to combine into one co-added image is denoted by $N_{im}$. In this work, we use $N_{im} = 2000$, if not specified otherwise; guidance on how to choose $N_{im}$ is provided in Sec.~\ref{subsec:choose}. The imaging field-of-view diameter is 9 arcseconds and each pixel is 0.021 arcsec by 0.021 arcsec.   To visualize the detector's performance, we use  Monte Carlo to calculate the probability density functions (PDF) of the photon counts for different ground-truth photon flux in a co-added image from 2000 PC images, shown in Fig.~\ref{fig:pdf}(a). Four of the PDF's are plotted in Fig.~\ref{fig:pdf}(b); the parameters in the simulations are listed in Table~\ref{table:parameter}. A photon rate of 0.1 photon/s (our expected maximum rate) is well within the linear response regime.

\begin{figure}
\begin{center}
\begin{tabular}{c}
\includegraphics[width=0.88 \textwidth,trim= 1.5cm 2.5cm 1cm  1.5cm ,clip]{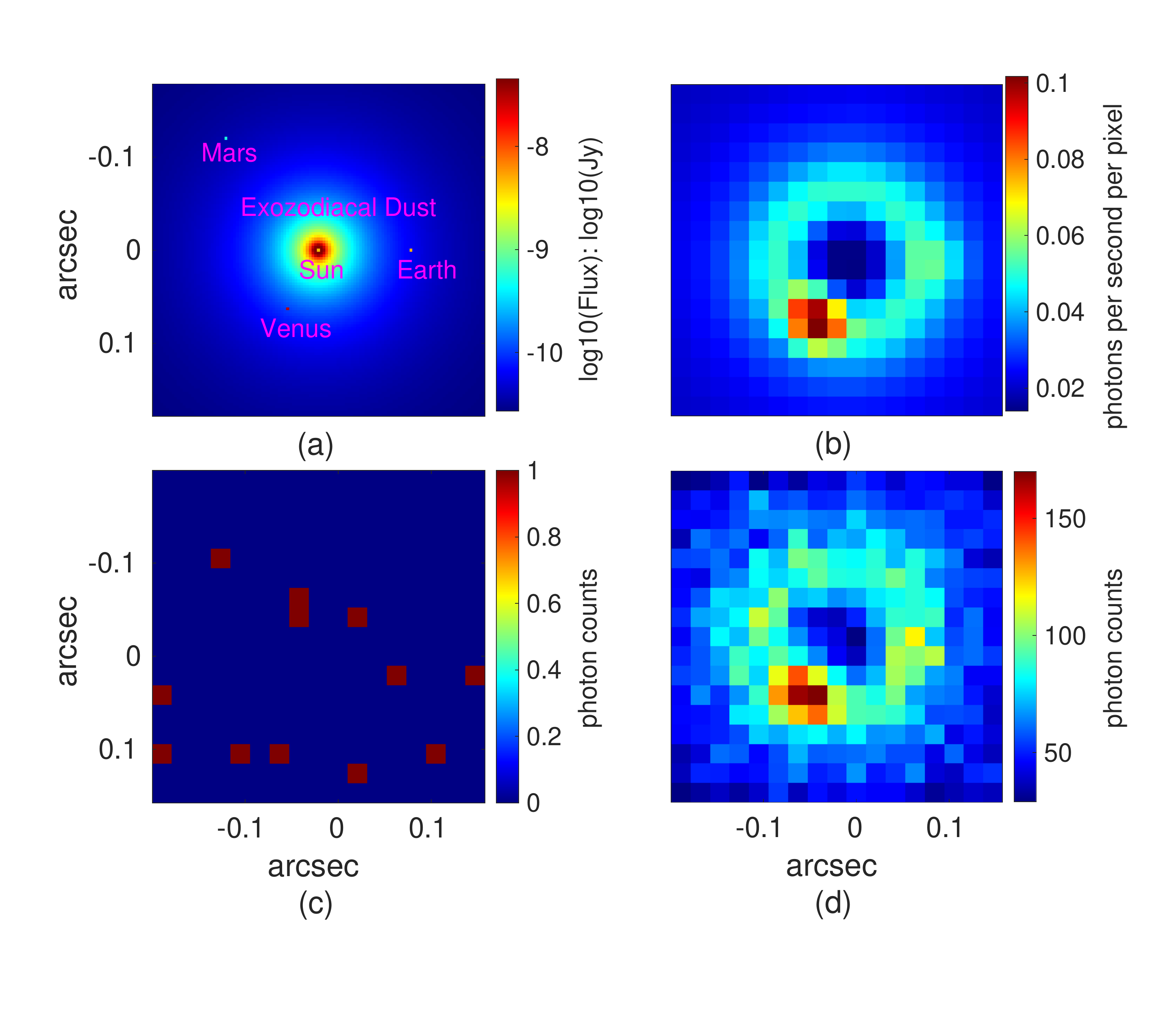}
\end{tabular}
\end{center}
\caption{The solar system as viewed from 10 pc $-$ and  its results from different stages in our simulation. They are zoomed-in views of the images  (a) is the original astronomical scene  from Haystacks project\cite{haystack}  at $\lambda=0.55~\upmu$m in   log10 scale flux (the central star is made $10^{10}$ times dimmer to reveal fainter features).  (b) is a simulated  result before including the detector noise, with a perfect starshade at $\lambda=0.55~\upmu$m and with a bandwidth of $\lambda=0.12~\upmu$m.  (c)  is one PC image. (d) is the co-added image from 2000 PC images. \label{fig:haystack}}
\end{figure}

Fig.~\ref{fig:haystack} shows one wavelength slice of a Haystacks  model, along with a few  results from different stages in our simulation.    Fig.~\ref{fig:haystack} (a) is the input of our simulation,  a discretized astrophysical scene of our solar system as viewed from 10 pc. 
The pixel size is small enough to make sure that the Haystacks models are high-fidelity spatially. The Haystacks scene includes the star, planets, interplanetary dust, and astrophysical background sources. As described in Fig.~\ref{fig:generation}, the light sources, i.e. Fig.~\ref{fig:haystack}(a), are diffracted at the starshade plane and then propagated to the telescope's pupil plane. A simple (aberration-free) telescope model propagates the pupil plane to the detector plane, the result of which is shown in Fig.~\ref{fig:haystack}(b). Then,  Fig.~\ref{fig:haystack}(b) is processed with the detector model to generate one single PC image, which is shown in Fig.~\ref{fig:haystack}(c). By combing 2000 PC images, we get the co-added image Fig.~\ref{fig:haystack}(d). The dark region in the center is the starlight suppression effect of the starshade. The brighter ring is the exozodiacal dust outside the starshade's IWA. Mars is too dim to be seen and Earth is of similar brightness as the exozodiacal dust. Venus is bright enough to be seen.  In the following sections, we propose a detection and characterization method of planetary signals for co-added images at $\lambda=0.55~\upmu$m and with a bandwidth of $\lambda=0.12~\upmu$m.

\begin{table}[ht!]
\centering
\caption{ Parameters for simulation of solar system as viewed from 10 parsec.}
\begin{tabular}{ |p{7.7cm}|p{2cm}|p{3.5cm}|  }
 \hline
 Parameter  & Value & Unit\\
 \hline
Sun flux& 45.66& $Jy$\\
Venus flux&  2.99e-8& $Jy$\\
Earth flux&  4.85e-9& $Jy$\\
Venus angular separation&  70& $mas$\\
Earth angular separation&  96& $mas$\\
Radius of the ROI & 100 & $mas$\\
Wavelength &0.55 &$~\mu m$\\
Bandwidth &0.12&$~\mu m$\\
Radius of the starshade  & 13 & $m$\\
Separation between starshade and telescope & 37.2e6 & $m$\\
Telescope diameter& 2.4& $m$\\
Detector's pixel size &0.021&$arcsec$\\
 Quantum efficiency  & 1 & $ph/e^-$\\
 Integration time  & 1 & $s$\\
 Clock-induced charge      &  0.01 & $e^-pixel^{-1}frame^{-1}$\\
 Dark current   & $2\times10^{-4}$ & $e^-pixel^{-1}s^{-1}$\\
 Electron-multiplying gain  & 2500 & $-$ \\
 PC Bias  & 200 & $e^-pixel^{-1}frame^{-1}$\\
 Standard deviation of readout noise & 100 & $e^-pixel^{-1}frame^{-1}$\\
 Threshold Parameter  & 5.5 & $-$ \\
 Number of PC image for one co-added image&2000& $-$ \\
 \hline
\end{tabular}
\label{table:parameter}
\end{table}

\section{Generalized Likelihood Ratio Test} \label{sec:glrt}

This section presents the GLRT as an automatic detection method for starshade missions.  Our work is motivated by the lack of any previous investigation into signal detection in starshade images. We begin the section by  reviewing past work on signal detection for direct imaging, most of which are specialized to  coronagraphic  images.

 The biggest challenge for  coronagraphic  images is the noise floor set by quasi-static speckles. Different observing techniques and post-processing methods have been developed to try to attenuate the speckles before attempting detection.   They are all based on differential imaging, which consists of estimating the star-only coronagraphic image and subtracting it from the science images (also called speckle subtraction technique). Differential imaging relies on specific observation strategies such as angular diversity (ADI)\cite{adi}, spectral diversity (SDI)\cite{ssdi}, or multiple reference star images (RSDI)\cite{rsdi} to generate the differential signal.   The various speckle subtraction methods developed for coronagraphic images may serve to improve detection in starshade images, but it is beyond the scope of this work to include them. We focus on planet detection in images that have not been post-processed and leave that work to be done in the future.

 In our work, we do not post-process the images    (as in speckle subtraction),  so we now move to detection methods. Ref.~\citenum{hyp} and Ref.~\citenum{linear} use hypothesis testing for the detection. The unknown parameters such as the planets' positions and intensities can be removed by marginalizing the probability\cite{hyp} or using worst-case values \cite{linear}. However, the choice of priors or using the worst-case values is an open question. Ref.~\citenum{linear} assumes a known constant background; in our method we will estimate the background with a maximum likelihood. Ref.~\citenum{Mawet14}    (SNRt map)  essentially tests if the intensity of a single test resolution is different from the other resolution elements in a $1\lambda/D$-wide annulus at the same radius. They use small sample statistics to address the problem of the statistical significance of detection when few realizations of spatial speckles vs azimuth are present. Ref.~\citenum{Cantalloube15}    (ANDROMEDA)   makes a signal template considering over/self subtraction caused by ADI rather than directly using a theoretical PSF template when calculating MLE. Ref.~\citenum{Ruffio17}   (FMMF)  also uses SNR based methods. They include KLIP-induced distortion in the match-filter template to estimate the signal intensity. The standard deviation is calculated at each pixel while masking a disk with a 5-pixel radius centered on that pixel from the annulus to prevent a planet biasing its own SNR. Ref.~\citenum{Flasseur18}    (PACO)  uses GLRT and uses the full-covariance rather than assume independent and identically distributed in the time dimension for the different frames. However, they assume the covariance is the same under $H0$, $H1$ and thus calculate only one covariance. Moreover, this covariance is not derived from the gaussian equation together with the signal intensity, and thus the result is not guaranteed to be MLE. Ref.~\citenum{Pairet19}     (STIM map) proposes STIM map, which uses a Modified Rician distribution, which is a heavy tail distribution compared to Gaussian, to model the speckles. Ref.~\citenum{Dahlqvist20}     (RSM map) uses a Markov process to model the same pixel throughout the different images. They claim the residual quasi-static speckles after ADI can be characterized by their mean and variance. Recent studies also take the detection problem as a binary classification problem, using random-forest classifiers  (SODIRF) and deep neural networks  (SODINN) \cite{sl}. However, these machine learning methods need very large training sets, which are difficult to generate for unknown planet signals, and  require a heuristic tuning of hyper-parameters.
 
 In this study, we also take the detection problem as a ``hypothesis testing'' problem and use a generalized likelihood ratio test. The null hypothesis  ($H_0$) is that there is no planet; the alternative hypothesis  ($H_1$) is that there is a planet. We compare the posterior probabilities under the two hypotheses to decide whether to reject the null. Instead of marginalization, we use MLE to first estimate the unknown parameters  (intensity, position and standard deviation of the noise) and then use them to calculate the likelihood ratio. Using this ratio, even when the maximum likelihood under the alternative hypothesis is low, we may still have a strong detection as long as the ratio is high (i.e., the pattern is much less likely from pure noise). We first introduced the GLRT model and its preliminary results in Ref.~\citenum{glrt}. In this section, we briefly describe the method, show recent improvements, and present results.

\subsection{GLRT model for the whole image}
The model for an image containing multiple planet signals, background, and noise and is given by
\begin{equation}
\label{model_im}
\bm{I}= \sum_{i =1}^{N_x}\sum_{j =1}^{N_y}[ \alpha_{i,j}  \bm{P}_{i,j}] + \bm{ b} + \bm{\omega}\,,
\end{equation}
where $\bm{I}$ is the matrix for an image; $ \alpha_{i,j}$ is the intensity of signals at pixel $(i,j)$ in unit of Jy (the value is zero if there is no signal at $(i,j)$); $ \bm{P}_{i,j}$ is the matrix denoting the values of a normalized PSF centered at $(i,j)$; $ \bm{b} $ is the matrix for background which contains star residual light and bias from the detector; $\bm{\omega}$ is the matrix for noise; $N_x,N_y$ are the number of pixels in x and y axis. 

We assume that the noise in different pixels is an independent and identically distributed Gaussian random variable with mean zero and unknown constant variance.  As each final image is the combination of many PC images,   a Gaussian distribution should be a good approximation for the noise due to the central limit theorem. As a Gaussian distribution is used as an approximation for the true underlying distribution, the estimation is called Gaussian quasi-maximum likelihood estimation (QMLE). As long as the quasi-likelihood function is not oversimplified, the QMLE is consistent and asymptotically normal. It is less efficient than the MLE, but may only be slightly less efficient if the quasi-likelihood is constructed so as to minimize the loss of information relative to the actual likelihood \cite{qmle}. We can then calculate the  QMLE of $\alpha$ and $\bm{b}$, which is equivalent to solving the optimization problem:
\begin{equation}
\label{opt}
\smash{\displaystyle\min_{\bm{\alpha}, \bm{b} }} ||\bm{I} - \sum_{i =1}^{N_x}\sum_{j =1}^{N_y}[ \alpha_{i,j}   \bm{P}_{i,j}] -\bm{ b} ||_2\,.
\end{equation}

This is an under-determined problem as we have $2N_xN_y$  unknown parameters, i.e., $\bm{\alpha,b}$  and only $N_xN_y$ data points, i.e., all the pixel values. Even assuming that we have a constant background in the whole image, we still have $N_xN_y+1$  unknown parameters. Moreover, the assumption of constant background is not ideal as the background may have local features. To approach the problem, we use the idea of divide-and-conquer.  If we assume no overlapping signals in the image, we are able to individually analyze smaller search areas that are the size of the PSF core.

\subsection{GLRT model for a search area}
\label{sec:area_model}
In a small search area with the size of the PSF core,  we can assume that there is only one planet signal. We also assume the background is constant, which should be a reasonable assumption over a small area. Moreover, noises are independent and identically distributed Gaussian random variables. Thus, the model for a search area is:
\begin{equation}
\label{model_pa}
\bm{x}_{i,j}= \alpha_{i,j}  \bm{P}_{i,j} + b_{i,j} \bm{1} + \bm{\omega }_{i,j}\,,
\end{equation}
where $\bm{x}_{i,j}$ denotes the search area centered on the global pixel (i,j); $ b_{i,j}$ is the background intensity; $\bm{\omega }_{i,j}$ denotes the Gaussian noise in this area. We only use the  central core of the PSF for $\bm{P}_{i,j}$ centered at (i,j).  The model states that the area we are observing,  $\bm{x}_{i,j}$, contains a signal centered at the center of this area along with constant background light and Gaussian noise. If there is no planet signal, $\alpha_{i,j}$ should be zero.
We stack pixel values in this target area into a column vector for easier mathematical manipulation. We do the same to the reference PSF and the noise matrix.  The local model can be reformulated as a classical linear model:
\begin{equation}
\label{H}
\bm{G_{i,j}}= \begin{bmatrix} P_{i,j}(1) & 1 \\ P_{i,j}(2) & 1 \\ \vdots & \vdots \\  P_{i,j}(N) & 1  \end{bmatrix} , \bm{\theta}_{i,j}=  \begin{bmatrix} \alpha_{i,j} \\  b_{i,j} \end{bmatrix} 
\end{equation}

\begin{equation}
\label{model}
\bm{x}_{i,j}= \bm{G}_{i,j} \bm{\theta}_{i,j} + \bm{w}_{i,j}\,,
\end{equation}
where $\bm{x}_{i,j}$ is the vectorized target area centered at $(i,j)$; $N$ is the number of pixels in the area; $b_{i,j}$ is the constant background; ${P}_{i,j}(m)$ is the value at the $m^\text{th}$ pixel in the known vectorized template PSF centered at $(i,j)$ for  a point source signal of normalized intensity; $\alpha_{i,j}$ is the planet's intensity, and $ \bm{w}_{i,j}$ is an $N \times 1$ noise vector.  $\bm{\theta}_{i,j}$ is unknown.

  The conditional probability of this search area can be written as
\begin{equation} \label{eq:prob_G}
\begin{aligned}
    L(\bm{\theta_{i,j}},\sigma_{i,j}^2) &    = p(\bm{x}_{i,j} |\bm{\theta_{i,j}},\sigma^2_{i,j}) \\
&    = \frac{1}{(2\pi \sigma^2_{i,j})^{N/2}} exp\left( -\frac{1}{2\sigma^2_{i,j}}  \left|\left|\bm{x}_{i,j} - \bm{G} \bm{\theta_{i,j}}\right|\right|_2  \right)  \,. 
\end{aligned}
\end{equation} 
  As the data $\bm{x}_{i,j}$ are known and the parameters $\bm{\theta_{i,j}}, \sigma_{i,j}^2$ are unknown, this probability function is a likelihood function for the unknown parameters (so it is also denoted as $L(\bm{\theta_{i,j}},\sigma_{i,j}^2)$ above). Taking the natural logarithm of both sides of Eq.~(\ref{eq:prob_G}), the log-likelihood of the search area in the co-added image is
\begin{equation} \label{eq:likelihood_G}
\begin{aligned}
  l(\bm{\theta_{i,j}},\sigma_{i,j}^2) &   = ln(L(\bm{\theta_{i,j}},\sigma_{i,j}^2))\\
&    =   -\frac{N}{2}ln(2\pi) -  \frac{N}{2}ln(\sigma_{i,j}^2)  -\frac{1}{2\sigma_{i,j}^2}  \left|\left|\bm{x}_{i,j} - \bm{G} \bm{\theta_{i,j}}\right|\right|_2 \,.
\end{aligned}
\end{equation}
  We maximize the log-likelihood function (equivalently maximizing the likelihood) to find the Maximum Likelihood estimator (the subscripts $i,j$ for the estimated parameters are left out for simplicity):
\begin{equation}
\label{equ:ML}
     (\hat{ \bm{\theta}}_{1},\hat{\sigma}^2_{1}) = \underset{\bm{\theta_{i,j}},\, \sigma^2_{i,j} }{\mathrm{argmax}} \; l(\bm{\theta_{i,j}},\sigma_{i,j}^2) \,.
\end{equation}
 As mentioned previously, a Gaussian distribution is used as an approximation for the true underlying distribution; the estimation is also QMLE. The resulting estimation under $H_1$ is:
\begin{equation}
\label{opt_area}
 \hat{ \bm{\theta}}_{1} =(\bm{G}_{i,j}^T\bm{G}_{i,j})^{-1}\bm{G}_{i,j}^T\bm{x}_{i,j}\,.
\end{equation}
   And the estimated variance under $H_1$ is:
\begin{equation}
\label{opt_area2}
     \hat{\sigma_1^2} = \frac{1}{N} (\bm{x}_{i,j} - \bm{G}_{i,j} \hat{ \bm{\theta}}_{1})^T (\bm{x}_{i,j}- \bm{G}_{i,j}  \hat{ \bm{\theta}}_{1}) \,.
\end{equation}
    Meanwhile, the QMLE under $H_0$ is:
\begin{equation}
\label{opt_area0}
      \hat{ \bm{\theta}}_{0} =   \hat{ \bm{\theta}}_{1} -   (\bm{G}_{i,j}^T\bm{G}_{i,j})^{-1}\bm{A}^T\left[\bm{A}(\bm{G}_{i,j}^T\bm{G}_{i,j})^{-1}\bm{A}^T\right]^{-1}(\bm{A} \hat{ \bm{\theta}}_{1})\,,
\end{equation}
  which is obtained from solving the constrained optimization problem:
\begin{equation}
\label{opt_area0_1}
\begin{aligned}
    \smash{\displaystyle\min_{\bm{\theta } }} &    \left|\left|\bm{x}_{i,j} - \bm{G}_{i,j} \bm{\theta}\right|\right|_2\\
     \textrm{s.t. } &    \bm{A} \bm{\theta} =0\,,
\end{aligned}
\end{equation}
 where $\bm{A}= [1, 0]$. And the estimated variance under $H_0$ is:
\begin{equation}
\label{opt_area3}
  \hat{\sigma_0^2} = \frac{1}{N} (\bm{x}_{i,j} - \bm{G}_{i,j}  \hat{ \bm{\theta}}_{0})^T (\bm{x}_{i,j} - \bm{G}  \hat{ \bm{\theta}}_{0})\,.
\end{equation}

 In the problem of parameter estimation, we obtain information about the unknown parameters from the observed data of the random variables from the probability distribution governed by the parameters. The Fisher information matrix is a way to quantify the amount of information that the observable random variables carry about the unknown parameters. The definition of the Fisher information matrix is 
\begin{equation} \label{eq:FIM}
\begin{aligned}
 \bm{I}(\bm{\phi})  &  =\mathrm{Var}_{\bm{\phi}} \{\nabla l(\bm{\phi})\}\\
&  = - E_{\bm{\phi}} \{\nabla^2 l(\bm{\phi})\} \,,
\end{aligned}
\end{equation}
 where the notation ``Var'' means the variance; ``E'' means expectation;  $\bm{\phi}$ is the vector of unknown parameters and $\bm{\phi}=\left(\alpha_{i,j}~~b_{i,j}~~\sigma^2_1\right)^T$ for the alternative hypothesis $H_1$. In our case (a linear model), the Fisher information matrix reduces to \cite{FIM_linear}
\begin{equation}
\label{eq:FIM_G}
 \bm{I}(\bm{\phi}) = \begin{pmatrix}
   -E\left( \frac{\partial^2 \,  l }{ \partial \, \bm{\theta}_{1}\, \partial \, \bm{\theta}_{1}^T}  \right) &  0\\ 
   0 &    -E\left( \frac{\partial^2  \, l }{ \partial \, (\sigma^2_1)^2 }  \right)
\end{pmatrix}
\,.
\end{equation}
 Due to the block structure of $\bm{I}(\bm{\phi}) $, the variance-covariance of $\hat{\bm{\theta}}_{G,1}$ can be estimated by $\bm{I}^{-1}_{\bm{\theta}_{G,1}}$ \cite{FIM_linear} where
\begin{equation}
    \label{equ:FIM_G_theta}
\begin{aligned}
     \bm{I}_{\hat{\bm{\theta}}_{1}} &  = - \frac{\partial^2 \,  l(\hat{\bm{\theta}}_{1}, \hat{\sigma}^2_{1})  }{ \partial \, \bm{\theta}_{1}\, \partial \, \bm{\theta}_{1}^T}   \\
    &  = \frac{G_{i,j}^TG_{i,j}}{\hat{\sigma}_1^2} \,,
\end{aligned}    
\end{equation}
 and we can also derive the confidence intervals for the QMLE\cite{fisherIn}: 
\begin{equation} \label{eq:alphaCI_G}
 \hat{\alpha} \pm z \sqrt{ (     \bm{I}_{\hat{\bm{\theta}}_{1}}^{-1}  )_{11} },
\end{equation}
 and
\begin{equation} \label{eq:betaCI_G}
  \hat{b} \pm z \sqrt{ (     \bm{I}_{\hat{\bm{\theta}}_{1}}^{-1}  )_{22} }, 
\end{equation}
 where $z$ is the appropriate  critical value (for example, 1.96 for 95 $\%$ confidence), and the notation $( \bm{I}_{\hat{\bm{\theta}}_{1}}^{-1})_{ii}$ means that we invert the matrix $\bm{I}_{\hat{\bm{\theta}}_{1}}$ first, and then take the $ii$ component of the inverted matrix. The variance of $\hat{\sigma}_1^2$ is estimated by  $\bm{I}^{-1}(\hat{\sigma}_1^2)$ \cite{FIM_linear} where
\begin{equation}
    \label{equ:FIM_G_sigma}
\begin{aligned}
     \bm{I}(\hat{\sigma}_1^2) &  = - \frac{\partial^2 \,  l(\hat{\bm{\theta}}_{1}, \hat{\sigma}^2_{1})  }{ \partial \, (\sigma^2_1)^2}   \\
    & = \frac{N}{2\hat{\sigma}_1^4} \,.
\end{aligned}    
\end{equation}

\subsection{Detection in a search area}
\label{sec:area_detection}
The detection problem becomes a hypothesis testing problem:
\begin{equation}
\label{Hyp}
\begin{split}
&H_0 :\bm{A} \bm{\theta}_{i,j}=\alpha_{i,j}  =0 \\  
&\mathrm{versus}       \\
&H_1 :\bm{A} \bm{\theta}_{i,j}=\alpha_{i,j}   \neq 0\,,
\end{split}
\end{equation}
where $\bm{A}= [1, 0]$. 

To decide which hypothesis is true, it is intuitive to compare the posterior probability of $H_1$ and  $H_0$ when $\bm{x}_{i,j} $ occurs. We use Bayes' rule and define the odds ratio:
\begin{equation}
\label{lr}
O(\bm{x}_{i,j} ) = \frac{P( H_1|\bm{x}_{i,j}  )}{P( H_0|\bm{x}_{i,j} )}= \frac{ \int f(z) P(\bm{x}_{i,j} |\alpha_{i,j}=z, b_{i,j},\sigma_{i,j}) dz} {P(\alpha_{i,j}=0)P(\bm{x}_{i,j} |\alpha_{i,j}=0, b_{i,j},\sigma_{i,j}),}\,,
\end{equation}
where $f(z)$ is the probability density function of the planet intensity   and $\sigma_{i,j}$ is the standard deviation of the noise.  We should decide $H_1$ is true if the odds ratio  is larger than one \cite{hyp}. However, we do not know $f(z)$,  $b_{i,j}$, or $\sigma_{i,j}$. It is reasonable to compare
\begin{equation}
\label{lr2}
  R (\bm{x}_{i,j}) = \frac{\displaystyle\max_{\alpha_{i,j}, b_{i,j}  ,\sigma_{i,j}}P(\bm{x}_{i,j}|\alpha_{i,j}, b_{i,j} ,\sigma_{i,j})}{\displaystyle\max_{b_{i,j}  ,\sigma_{i,j}} P(\bm{x}_{i,j}|\alpha_{i,j}=0,  b_{i,j} ,\sigma_{i,j})} 
\end{equation}
to a threshold. This test is called the generalized likelihood ratio test  (GLRT). Moreover, not only is the planet intensity $\alpha_{i,j}$ unknown, so are the background intensity and noise statistics. We use the QMLE mentioned above. Then, the ratio becomes 
\begin{equation}
\label{lr3}
  R (\bm{x}_{i,j}) = \frac{P(\bm{x}_{i,j}| \hat{\bm{\theta}}_{i,j,1} , \hat{\sigma}_{i,j,1} ; H_1 )}{P(\bm{x}_{i,j}| \hat{\bm{\theta}}_{i,j,0}  , \hat{\sigma}_{i,j,0} ; H_0 )}\,,
\end{equation}
where $ \hat{\bm{\theta}}_{i,j,k}  , \hat{\sigma}_{i,j,k}$ are the estimated values under hypothesis $H_k$,  \{$k={0,1}$\}. 

To simplify further,  $H_1 $ is favored, i.e., a planet is more likely to exist at the test location, if
\begin{equation}
\label{T}
\begin{split}
 T(\bm{x}_{i,j})& = (N-2) (   R (\bm{x}_{i,j})^{\frac{2}{N}}-1)\\
&= (N-2) \frac{\hat{\sigma}^2_{i,j,0}-\hat{\sigma}^2_{i,j,1}}{\hat{\sigma}^2_{i,j,1}} \\
&= (N-2) \frac{  \hat{\bm{\theta}}_{i,j,1}^T \bm{A}^T[\bm{A} (\bm{G}_{i,j}^T \bm{G}_{i,j} )^{-1} \bm{A}^T ]^{-1} \bm{A} \hat{\bm{\theta}}_{i,j,1} }{ \bm{x}_{i,j}^T( \bm{1}_N -  \bm{G}_{i,j} (\bm{G}_{i,j}^T \bm{G}_{i,j} )^{-1} \bm{G}_{i,j}^T  )\bm{x}_{i,j} } > \gamma\,,
\end{split}
\end{equation}
where the threshold $\gamma$ is based on the detection performance  \cite{kay}. The probability of false alarms (false alarm rate) $P_{FA}$ and probability of detections $P_{D}$ are given by
\begin{equation}
\label{pfa}
P_{\bm{FA}  i,j}  = \int_{T(\bm{x}_{i,j}) > \gamma} p(\bm{x}_{i,j}|H_0 )\, d\bm{x}_{i,j}= Q_{F_{1,N-2}}(\gamma) 
\end{equation}
\begin{equation}
\label{pd}
P_{\bm{D}  i,j} = \int_{T(\bm{x}_{i,j}) > \gamma} p(\bm{x}_{i,j}|H_1 )\, d\bm{x}_{i,j}= Q_{F^\prime_{1,N-2}(\lambda_{i,j})}(\gamma)\,,
\end{equation}
where Q is the probability of exceeding a given value; $F_{1,N-2}$ is an F distribution with one numerator degree of freedom and N-2 denominator degrees of freedom; and $F^\prime_{1,N-2}(\lambda_{i,j})$ is a noncentral F distribution with one numerator degree of freedom and N-2 denominator degrees of freedom and noncentrality parameter $\lambda_{i,j}$ \cite{kay}. $\lambda_{i,j}$ is given by
\begin{equation}
\label{lam}
\begin{split}
& \lambda_{i,j} = \frac{  \bm{\theta}_{i,j,1}^T \bm{A}^T[\bm{A} (\bm{G}_{i,j}^T \bm{G}_{i,j} )^{-1} \bm{A}^T ]^{-1} \bm{A} \bm{\theta}_{i,j,1} }{\sigma^2}\,, 
\end{split}
\end{equation}
where $\bm{\theta}_{i,j}$ is the true value under $H_1$ and $\sigma^2$ is the true variance of the noise. This tells us that the probability of a false alarm only depends on the threshold, but the probability of detection depends on the planet intensity. The brighter the planet is, the higher the detection probability.

\subsection{Detection in the whole image}
We have built a model and corresponding detection and estimation method for a search area that is the size of the PSF core. However, the potential planets' locations are unknown, so to perform detection in the whole image, we will traverse the whole image using the method outlined in Sec.~\ref{sec:area_detection}. For each pixel, we use the detection area centered at this pixel, that is to say, we test $H_1$: there is a planet centered at this pixel, against $H_0$: there is only constant background there. After calculating for all the pixels in the image, we choose a false alarm rate and apply its corresponding threshold. Examples are illustrated in Fig.~\ref{fig:glrtr}.  As the PSF changes with the distance from the starshade, our detection algorithm uses a library of  reference PSFs.

When the pixel is at the boundary of the PSF in the image, shown as a  magenta asterisk in Fig.~\ref{fig:glrtr}(b), the detection area only contains part of the planet which is not centered at the pixel, so neither $H_1$ nor $H_0$ is true  and the MLE of planet intensity can be negative. Thus, we set those negative estimates   as zero and thus ${T(\bm{x}_{i,j})} =0 $. 

After thresholding, we get a binary image. Generally speaking, some pixels next to the signal center will also be detected. Thus, to estimate the position of the planet, we first find the convex hulls in the thresholded image and find the minimal circle bounding each convex hull. The center of the circles will be the estimates for the planets' positions. The estimated planet intensity is the MLE of $I$ at the estimated planet position. As we have shown in our previous work \cite{simulation}, the PSF changes with the distance away from the starshade center. If we use only one PSF template in GLRT, normally the one without a starshade, we can have a higher false alarm rate, worse position estimation, and worse intensity estimation. Thus, we need to have a library of PSFs at difference distances from the starshade center for the GLRT model. In this paper, we define $ \mathrm{Intensity~Error} = \frac{(\mathrm{Estimated~Intensity}-\mathrm{Real~Intensity})}{\mathrm{Real~Intensity}}$ and report its value for all cases. 

\subsection{Results}
\begin{figure}[ht!]
\begin{center}
\begin{tabular}{c}
\includegraphics[width=0.95\textwidth]{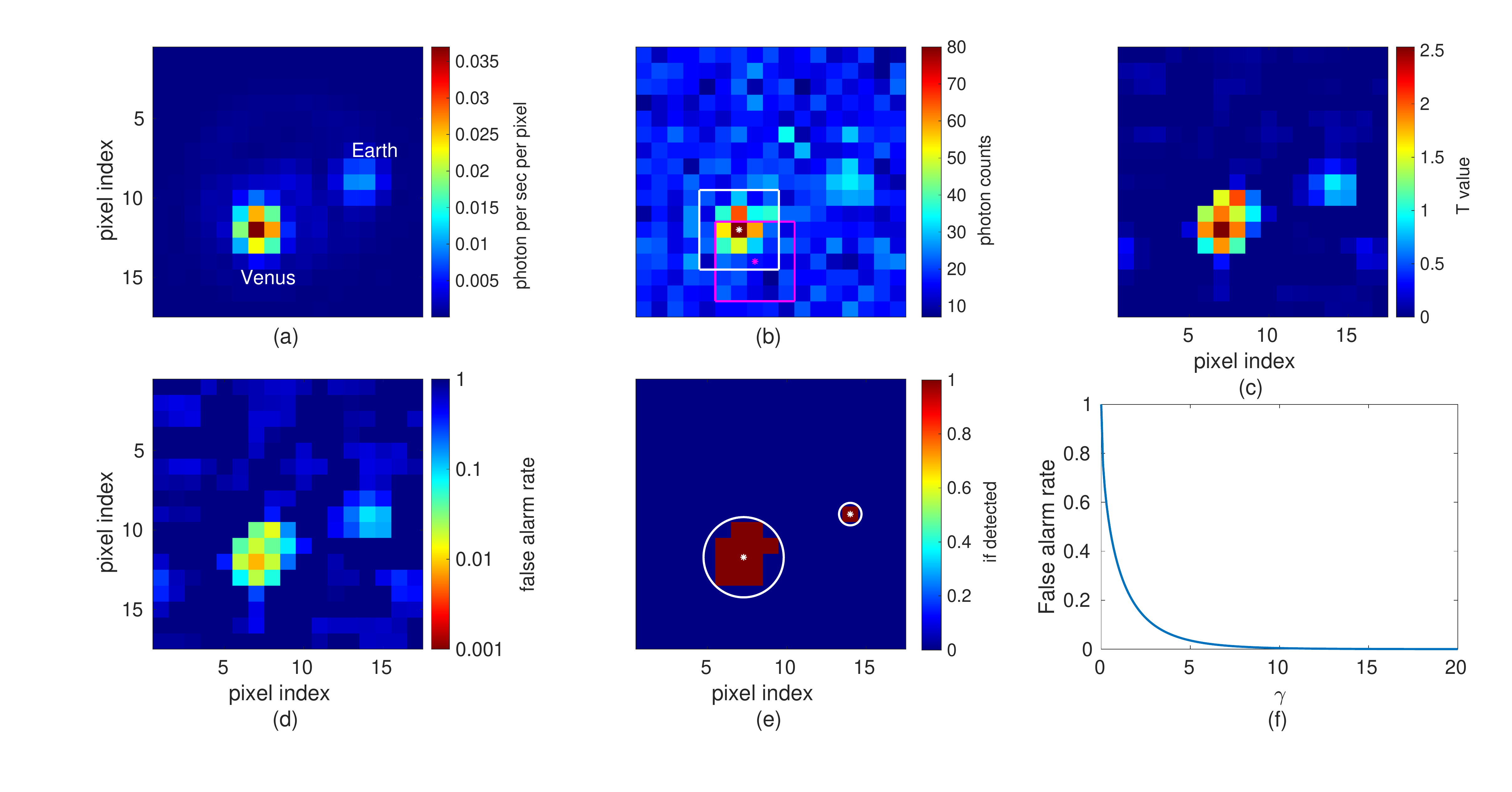}
\end{tabular}
\end{center}
\caption{Example of the GLRT detection on an image with perfect starshade and without exozodiacal dust. (a) Noiseless image. It should be take n as ground truth for image processing. (b) Image with detector noise. Examples of search areas are also shown. The white asterisk and box centered on pixel (7,12) form data $\bm{x}_{7,12}$ in Eq.~(\ref{model_pa}). The magenta box forms data $\bm{x}_{8,14}$ and is the case where the search area is at the edge of the PSF. (c) The T values from  Eq.~(\ref{T}) in each pixel. (d) False alarm rate map. (e) Binary detection image after thresholding. We apply a threshold of 0.7354 to the T map, which results in 0.4 false alarm rate. The white circles are the minimal bounding circles used to estimate the planet position. (f) The relationship between threshold and false alarm rate from  Eq.~(\ref{pfa}). \label{fig:glrtr}}
\end{figure}

\begin{figure}[ht!]
\begin{center}
\begin{tabular}{c}
\includegraphics[width=0.95\textwidth]{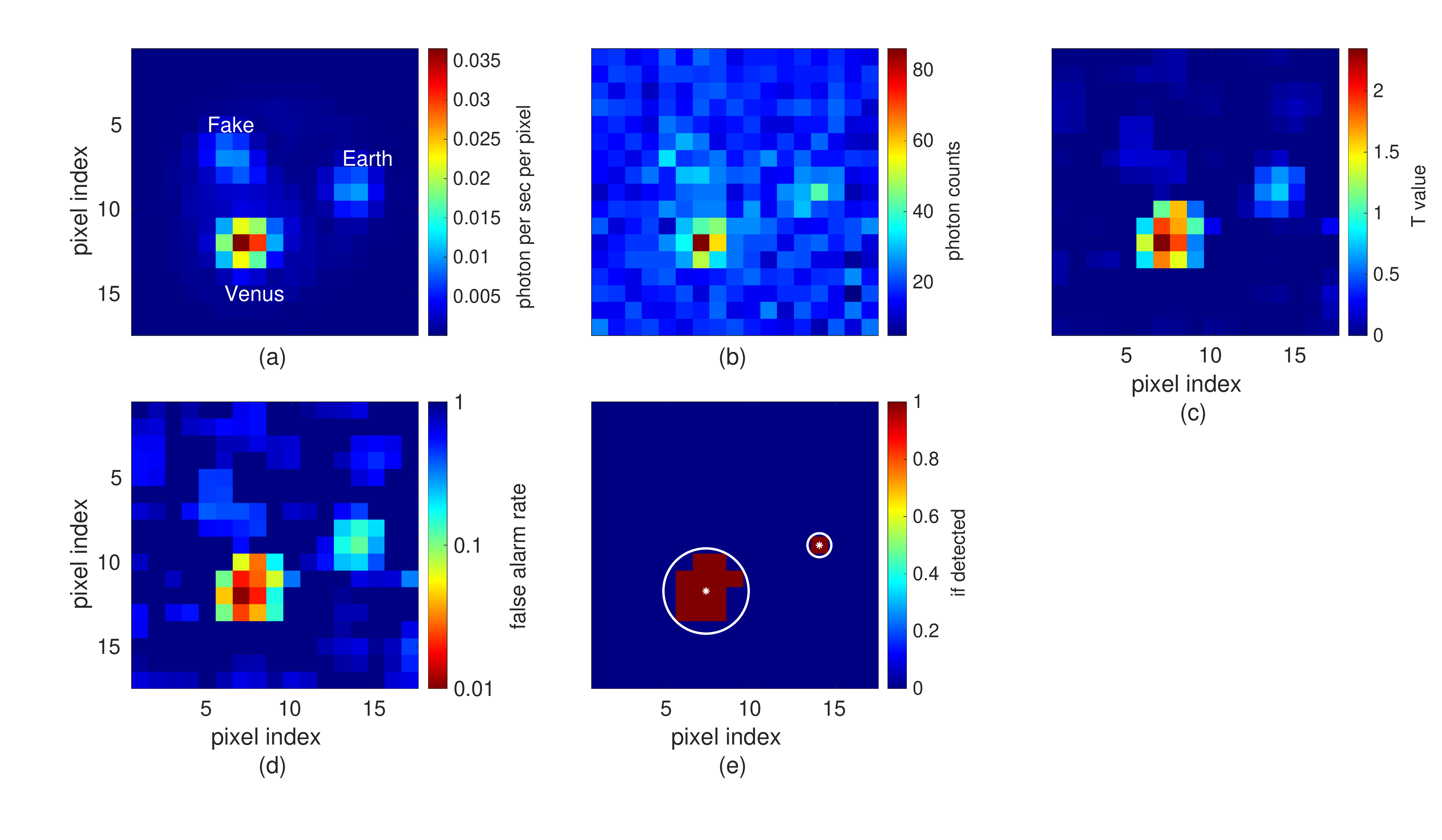}
\end{tabular}
\end{center}
\caption{Example of the GLRT detection on an image with clipped starshade. (a) Noiseless image. The defect on the starshade causes a bright spot that resembles a planet. (b) Image with detector noise applied. (c) The T values from  Eq.~(\ref{T}) in each pixel. (d) False alarm rate map. (e) Binary detection image after thresholding. We apply a threshold of 0.7354 to the T map, which results in 0.4 false alarm rate. GLRT successfully detects Venus and Earth and ignores the fake signal in the image. \label{fig:clip2}}
\end{figure}

Two detection and estimation examples are presented in Figs.~\ref{fig:glrtr} and \ref{fig:clip2}. In Fig.~\ref{fig:glrtr}, the detection process is demonstrated step by step with an image with perfect starshade. In Fig.~\ref{fig:clip2}, we demonstrate the GLRT's ability to distinguish real signals from fake ones by providing the detection result for an image with a starshade with truncated tips. Compared to the perfect starshade, the tip of one of this starshade's petals is truncated by 4.9 mm and resulting in a tip width of 24.2 mm. We refer to this starshade as the  clipped starshade. This defect causes a bright spot in the image, which could be mistaken for a planet. We still use the PSF templates of the perfect starshade for this clipped starshade case. GLRT successfully detects Venus and Earth from the fake signal. The errors of intensity and position estimation for these two examples are presented in Table.~\ref{tab:diffSS}. The fake signal is close to Venus, so the intensity estimation of Venus is degraded for the clipped starshade case. The pixel size of the images is 0.021 arcsec, so the position estimation is accurate to the pixel level.

\begin{table}[ht]
\caption{Intensity estimation error and position estimation error comparison between results in images using different starshades.} 
\label{tab:diffSS}
\begin{center}       
\begin{tabular}{|l|l|l|l|l|} 
\hline
\rule[-1ex]{0pt}{3.5ex} Starshade &  \begin{tabular}{@{}c@{}}Intensity Error \\ for Venus  \end{tabular}  &  \begin{tabular}{@{}c@{}}Intensity Error \\for Earth \end{tabular}  &\begin{tabular}{@{}c@{}} Position Error\\for Venus (arcsec) \end{tabular} &\begin{tabular}{@{}c@{}}Position Error\\for Earth (arcsec)\end{tabular}  \\
\hline
\rule[-1ex]{0pt}{3.5ex} Perfect starshade &	1.5\% &  1.2\% &	3E-03 &   9.5E-03 \\
\hline
\rule[-1ex]{0pt}{3.5ex} Clipped starshade &	20.5\%    &    4.1\% & 	3E-03 &   9.5E-03  \\
\hline
\end{tabular}
\end{center}
\end{table}

\begin{figure}
\centering
  \includegraphics[width=0.98\linewidth,trim= 2cm 0cm 2cm 0cm ,clip]{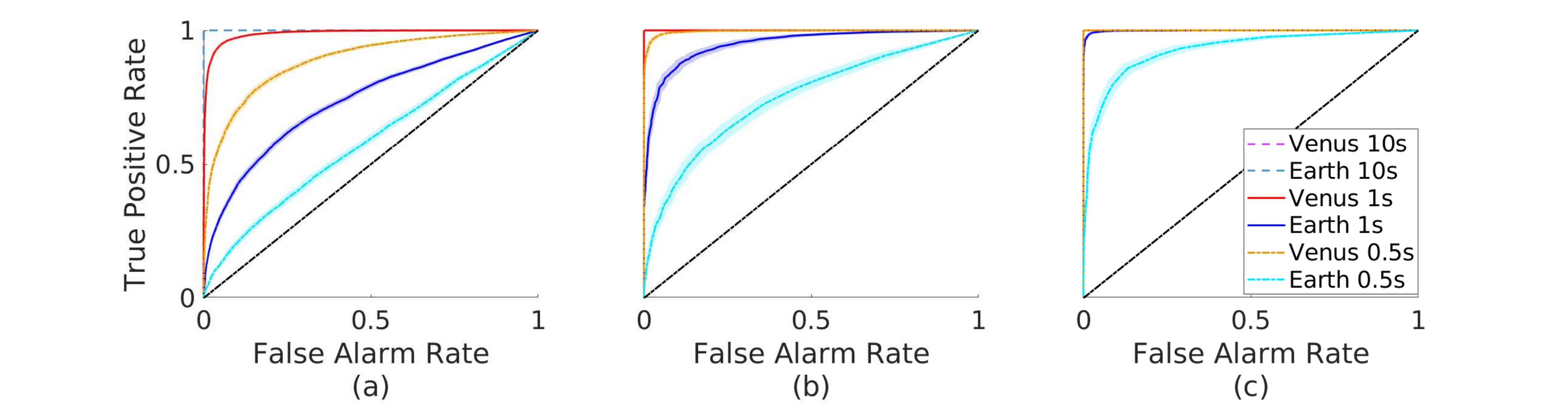}
\caption{  ROC with confidence interval for Venus and Earth using GLRT with different integration time. (a) ROC for co-added images each from $N_{im}=200$ PC images. The curves for an integration time of 10s overlaps with the plot's boundary, which indicates perfect detection performance.  (b) ROC for co-added images each from $N_{im}=700$ PC images. (c) ROC for co-added images each from $N_{im}=2000$ PC images. The shaded region behind each ROC curve is its 95$\%$ confidence interval.} 
\label{fig:roc}
\end{figure}

We also calculate the receiver operating characteristic (ROC) curves for Venus and Earth shown in Fig.~\ref{fig:roc}, where we compare the performance of GLRT for co-added images with different numbers of total images, which is denoted as $N_{im}$. Eq.~(\ref{pfa}) and Eq.~(\ref{pd}) give the theoretical false alarm rate and true positive rate under a Gaussian assumption, and are therefore only an approximation. Moreover, as shown in Eq.~(\ref{lam}), the calculation of the true positive rate needs the value of the true variance, which we need to estimate. Thus, to more accurately demonstrate the detection's performance, we use a Monte Carlo simulation. To calculate the ROC curves, we apply GLRT to get the false alarm rate map for a set of different thresholds and record if it results in a detection or missed detection of Earth, Venus and a background pixel. We run a large number of trials, which is denoted as $n_{trials}$, and record the ratio of detection of Earth and Venus as the true positive rate for Earth and Venus, and record the ratio of detection of the background pixel as the false positive rate. The confidence interval (CI) of a proportion $\hat{p}$ is \cite{statistics}:
\begin{equation}
\left(\hat{p} - z_{1-\alpha/2} \sqrt{ \frac{\hat{p}(1-\hat{p})}{n_{trials}} },~\hat{p} + z_{1-\alpha/2} \sqrt{ \frac{\hat{p}(1-\hat{p})}{n_{trials}} } \right).    
\label{eq:ci}
\end{equation}
Thus, for a point $(\hat{p}_{false}, \hat{p}_{positive})$ on a ROC curve, we take the confidence interval using the two points:
\begin{equation}
\left(\hat{p}_{false} - z_{1-\alpha/2} \sqrt{ \frac{\hat{p}_{false}(1-\hat{p}_{false})}{n_{trials}} },~\hat{p}_{positive} + z_{1-\alpha/2} \sqrt{ \frac{\hat{p}_{positive}(1-\hat{p}_{positive})}{n_{trials}} } \right)
\label{eq:cil}
\end{equation}
and
\begin{equation}
\left(\hat{p}_{false} + z_{1-\alpha/2} \sqrt{ \frac{\hat{p}_{false}(1-\hat{p}_{false})}{n_{trials}} },~\hat{p}_{positive} - z_{1-\alpha/2} \sqrt{ \frac{\hat{p}_{positive}(1-\hat{p}_{positive})}{n_{trials}} } \right).
\label{eq:cr}
\end{equation}  
For each ROC curve, we apply this two boundaries to calculate the shaded area as confidence interval. ROC curves for Venus and Earth with different numbers of PC images to combine into one co-added image $N_{im}$,  (with different integration times) are shown in Fig.~\ref{fig:roc}. As Venus is brighter than Earth, the performance for Venus is better than that for Earth.   Increasing the integration time has the similar effect as increasing the planet intensity because both increase the expected number of photons arriving on the pixel. Moreover, the performance for both Venus and Earth is better with the more number of PC images in the co-added image.

\subsection{Optimal number of PC images for one co-added image} \label{subsec:choose}

As we mentioned before, the number of PC images to combine into one co-added image $N_{im}$, is a important hyperparameter to be chosen before doing detection via most of the methods. 
For example, GLRT's performances varies with the number of PC image for one co-added image, as shown in Fig.~\ref{fig:roc}. With too small $N_{im}$, the true positive rate and false alarm rate for the detection may not be desirable. With too big $N_{im}$, precious observation time would be wasted. To choose the best $N_{im}$, we can utilize the ROC curves.  

\begin{figure}
\centering
  \includegraphics[width=0.7\linewidth]{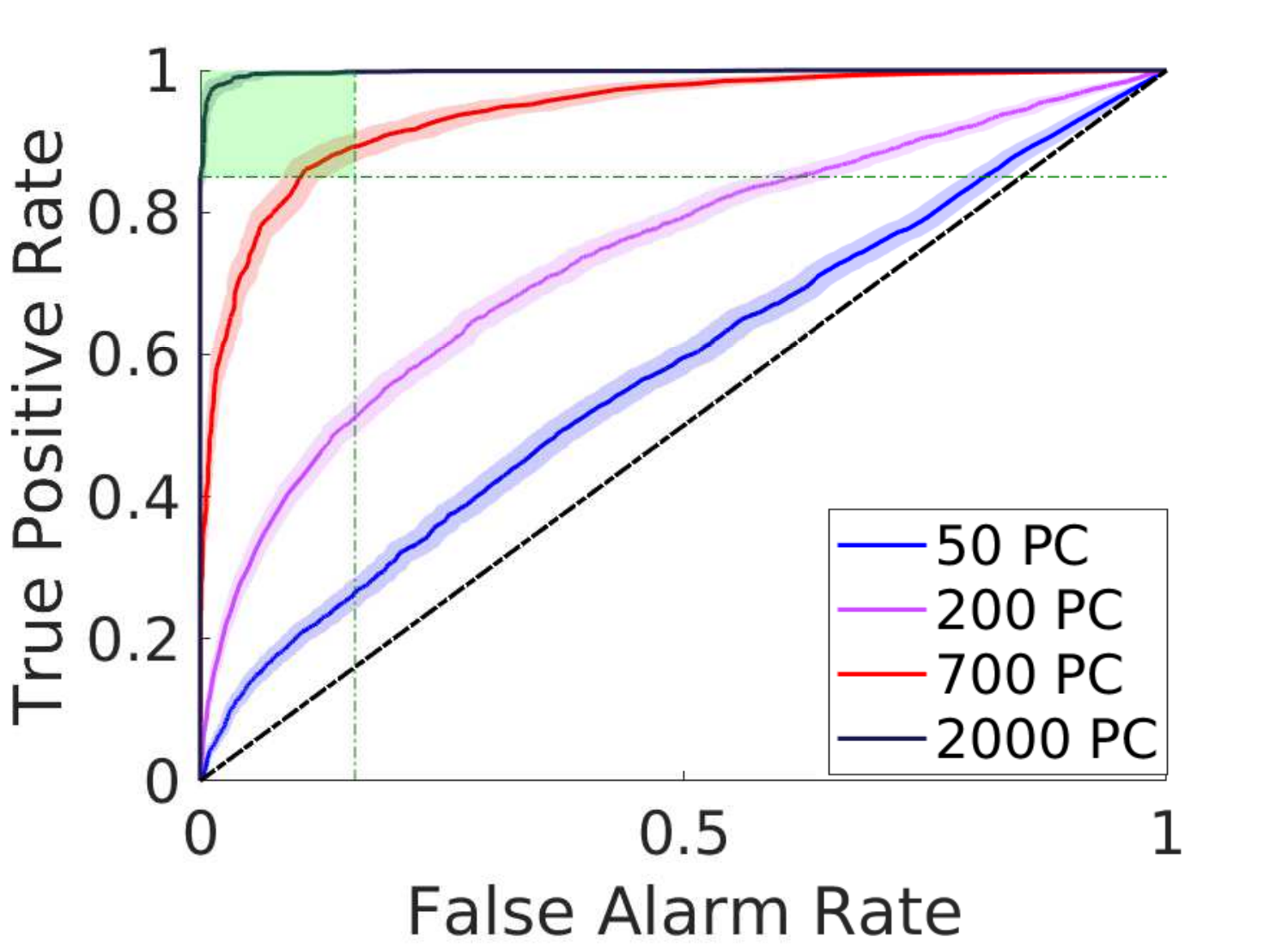}
\caption{ ROC with confidence interval for Earth using GLRT with different  $N_{im}$. The horizontal green dotted line is the acceptable true positive rate (TP) and the vertical one is the acceptable false alarm rate (FA) region. Thus, the acceptable true TP   and acceptable  FA region is the shaded green area. The ROC with $N_{im}=700$ is the first ROC reaching the acceptable region.  } 
\label{fig:stop}
\end{figure}

First,  given integration time, three parameters  need to be specified: the minimum planet intensity to be detected, the maximum false alarm rate that can be accepted, and the minimum true positive rate that is acceptable. Then, for a different $N_{im}$, the ROC is calculated via monte carlo simulation. Finally, the minimum  $N_{im}$ that can reach the requirements are chosen. For example, if we assume the dimmest planet has the same intensity as Earth, the maximum acceptable false alarm rate is 0.16 and the minimum acceptable true positive rate is 0.85. The acceptable false alarm rate, true positive rate pairs are in the shaded green region in Fig.~\ref{fig:stop}. We calculate the ROCs with different $N_{im}$ for integration time 1s and find that the ROC of $N_{im}=700$ is the first one to reach the green region in Fig.~\ref{fig:stop}. Thus we choose $N_{im}=700$ as the optimal number of PC images to be co-added into the final image.

\subsection{Comparison with other methods}

\begin{figure}[ht!]
\centering
\includegraphics[width=0.99 \textwidth,trim= 0  0  0 0cm ,clip]{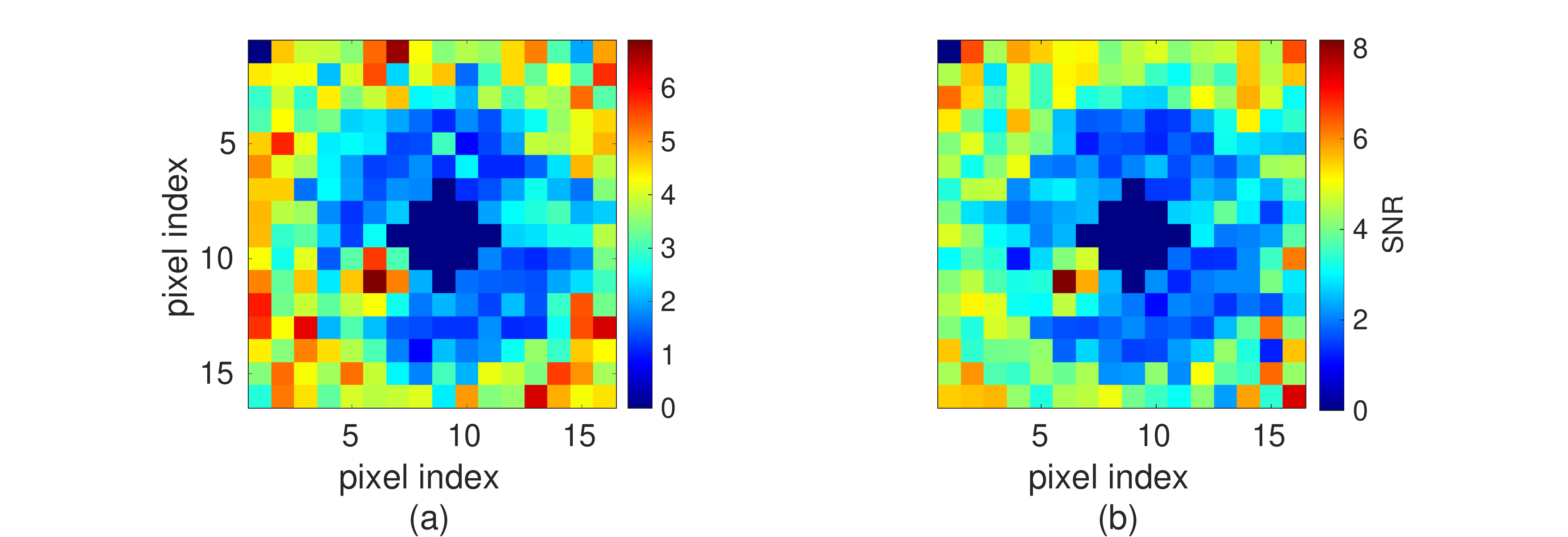}
\caption{ The SNR map   calculated with the pyKLIP package\cite{pypoint} (a) result for the image in Fig.~\ref{fig:glrtr}(b). (b) result for the image Fig.~\ref{fig:clip2}(b). }
\label{fig:pyklip}
\end{figure}

\begin{figure}
\centering
  \includegraphics[width=0.98\linewidth,trim= 2cm 0cm 2cm 0cm ,clip]{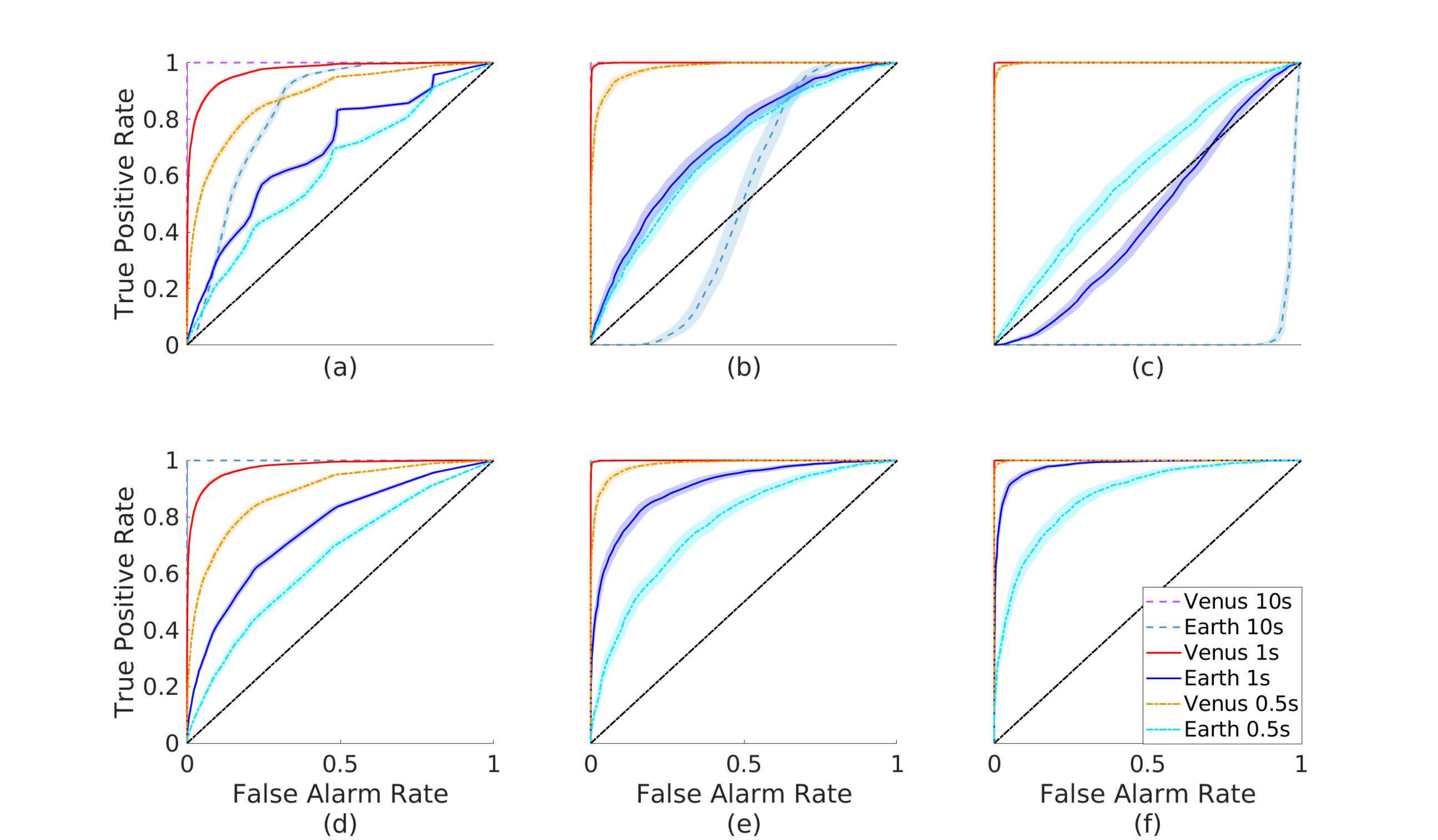}
\caption{  ROC with confidence interval for Venus and Earth using pyKLIP with different integration times. The first row is the result where Earth and Venus are at the real locations as shown in Fig.~\ref{fig:glrtr}(a). The second row is the result where we eliminate the impact of planets for each other: when calculating the ROCs for Earth, we didn't include Venus in the images;  when calculating the ROCs for Venus, we didn't include Earth in the images. The first column: ROC for co-added images each from $N_{im}=200$ PC images.  The second column: ROC from $N_{im}=700$ PC images. The third column: ROC from $N_{im}=2000$ PC images.  } 
\label{fig:pyklip_diff_intt}
\end{figure}

We also compared GLRT with the performance  of the detection method based on SNR map implemented in pyKLIP\cite{pyklip}. The algorithm computes the standard deviation in concentric annuli  after masking the signal area in question as the level of noise in SNR. The width of the annuli used is the diameter of the PSF core in this paper. The SNR maps for the three example co-added images in Fig.~\ref{fig:glrtr}(b) and Fig.~\ref{fig:clip2}(b)  are shown in  Fig.~\ref{fig:pyklip} to help visually compare the performance with the GLRT method.   ROC curves are also calculated, shown in Fig.~\ref{fig:pyklip_diff_intt}. The calculation uses the same set of images as the ones used for Fig.~\ref{fig:roc}. As it is hard to visually compare the curves, we list the area under the ROC curve (AUC) for all the curves in Table.~\ref{tab:auc}.  AUC is an aggregate measure of performance across all possible thresholds. It can be interpreted as the probability that the model ranks a random positive example higher than a random negative example. AUC is 1 if the model's decisions are all correct and 0 if all wrong. The disadvantage of this SNR definition is that the standard deviation calculation will be biased by the presence of point sources in annuli. In our case, the radius of Venus and Earth from the image center is close, so the signals are partially in each other's annuli for the calculation of the noise standard deviation. Thus, the SNR is biased, which is validated by the deteriorating performance for Earth when signals become stronger due to increased integration time or $N_{im}$, shown in Fig.~\ref{fig:pyklip_diff_intt}(b) and (c). Overall, GLRT outperforms SNR method on images that have not been post-processed; it is beyond the scope of this work to examine the effects of post-processing on each detection method.

\begin{table}[ht]
\caption{  Comparison of area under the curve (AUC) for GLRT and SNR method from pyKLIP\cite{pyklip}. Unimpacted means only one planet is included in the image simulation.}
\label{tab:auc}
\begin{center}       
\begin{tabular}{|l|l|l|l|l|l|l|} 
\hline
\rule[-1ex]{0pt}{3.5ex}   &  \begin{tabular}{@{}c@{}} Venus \\ 10s \end{tabular}  &  \begin{tabular}{@{}c@{}}Earth\\10s \end{tabular}  &\begin{tabular}{@{}c@{}} Venus \\1s\end{tabular} &\begin{tabular}{@{}c@{}}Earth\\1s\end{tabular}&\begin{tabular}{@{}c@{}} Venus \\0.5s\end{tabular} &\begin{tabular}{@{}c@{}}Earth\\0.5s\end{tabular}   \\
\hline
\rule[-1ex]{0pt}{3.5ex} GLRT, 200PC & 1  & 1 &	0.9883 &   0.7374 & 0.8880 & 0.5797  \\
\hline
\rule[-1ex]{0pt}{3.5ex} GLRT, 700PC &	1   &  1 & 1 &  0.9503 & 0.9963& 0.7490\\
\hline
\rule[-1ex]{0pt}{3.5ex} GLRT, 2000PC &	1 & 1 & 1 & 0.9987 & 1 &  0.9275  \\
\hline
\rule[-1ex]{0pt}{3.5ex} SNR, 200PC & 1  & 0.8299 &	0.9714 &   0.6953 & 0.8804 & 0.6219  \\
\hline
\rule[-1ex]{0pt}{3.5ex} SNR, 700PC &	1   &  0.5059  & 0.9993 &   0.7159& 0.9799& 0.6863\\
\hline
\rule[-1ex]{0pt}{3.5ex} SNR, 2000PC & 1  &  0.0325 & 1 &  0.4554 & 0.9985 & 0.6112  \\
\hline
\rule[-1ex]{0pt}{3.5ex} unimpacted SNR, 200PC &	1   &  0.9999 & 0.9768 &  0.7595 & 0.8870 & 0.6529   \\
\hline
\rule[-1ex]{0pt}{3.5ex} unimpacted SNR, 700PC &	 1  &   1  & 0.9997 & 0.9069 & 0.9848 & 0.7711    \\
\hline
\rule[-1ex]{0pt}{3.5ex} unimpacted SNR, 2000PC &1
&1&	1 &  0.9806&	0.9995 &   0.8753\\
\hline
\end{tabular}
\end{center}
\end{table}

\section{Iterative GLRT for Exozodiacal Dust} \label{subsec:dust}
Exozodiacal dust is debris in the habitable zones of stars believed to come from extrasolar asteroids and comets\cite{dust}. Though the true structure of exozodiacal dust is unknown, as a first attempt, we simply assume it is axisymmetric, which is believed to be a reasonable approximation for small dust grains \cite{sysmdust}. When the intensity of exozodiacal dust is similar to that of a target planet, the methods mentioned above have difficulty detecting the planet's signal. We develop here an iterative GLRT to tackle this problem. It is essentially an Expectation-Maximization (EM) algorithm\cite{EM}. The planets' signals and the exozodiacal dust are both unknown and need to be estimated. However, it is hard to estimate them accurately at the same time, and their estimation can influence each other. The solution is to iteratively estimate either the planets' signal or the exozodiacal dust first, and then use the estimation as a known factor to estimate the other until both estimates converge.

When there is exozodiacal dust, the model in Eq.~(\ref{model_im}), specified at pixel $(x,y)$, becomes:
\begin{equation}
\label{model_dust}
\bm{I}(x,y)= \sum_{i=1}^{N_x}\sum_{j=1}^{N_y}[\alpha_{i,j}  \bm{P}_{i,j}(x,y)] + \bm{ b}(x,y) + \bm{d}(x,y)+\bm{\omega}(x,y)\,,
\end{equation}
where $ \bm{d}(x,y)$ is the exozodiacal dust at pixel $(x,y)$ which is also unknown. We now have $3N_xN_y$ unknown parameters.

\begin{figure}[ht!]
\begin{center}
\begin{tabular}{c}
\includegraphics[width=\textwidth]{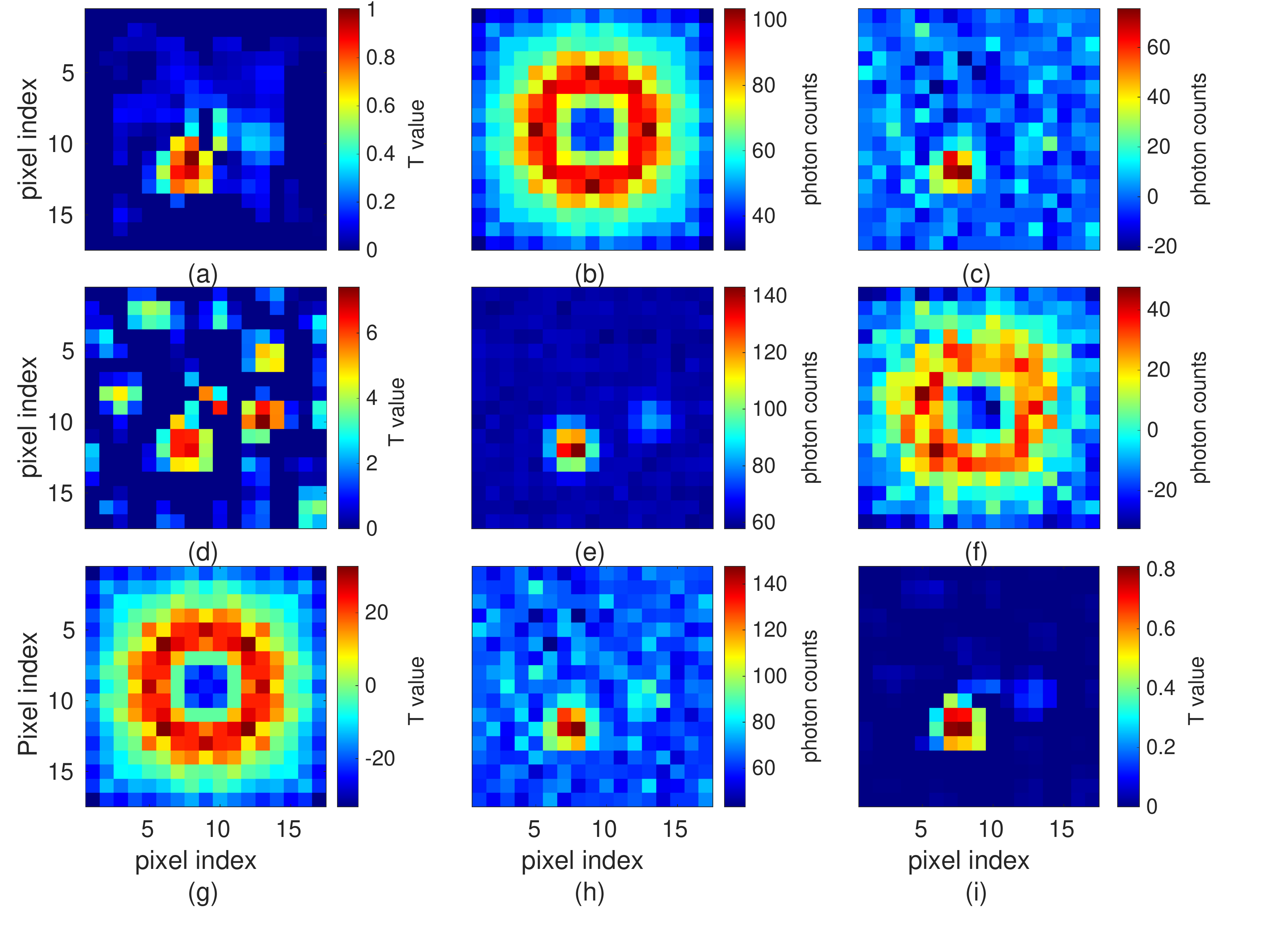}
\end{tabular}
\end{center}
\caption{Example of iterative GLRT applied to Fig.~\ref{fig:haystack}(f). (a)  T value map for Fig.~\ref{fig:haystack}(f). (b)  The median value  for each radius of Fig.~\ref{fig:haystack}(f), i.e., the exozodiacal dust estimation  $\bm{d }^{*}$ at initial step. (c) The residual $\bm{I_b}$ after subtracting  $\bm{d }^{*}$ from $\bm{I}$, i.e., the residual after subtracting (b)  from Fig.~\ref{fig:haystack}(f). It is the initial estimation for the underlying image with only planets. (d) T map for (c). (e) The new estimate of planets $\sum_{i=1}^{N_x}\sum_{j=1}^{N_y}[\alpha_{i,j}  \bm{P}_{i,j}(x,y)]$. After applying GLRT on (c) and get detection, we also obtained the intensity and position estimation of the planets. (f) Exozodiacal dust $I_p$ after subtracting estimated planet signals (e) from the original image Fig.~\ref{fig:haystack}(f). (g) The dust estimation at the final step. (h) The final residual $\bm{I_b}$, i.e., the residual after subtracting (g) from Fig.~\ref{fig:haystack}(f). It is the final estimation for the underlying image with only planets. (i) T value map for (h). \label{fig:FAR}}
\end{figure}

The exozodiacal dust degrades the detection and estimation. For example, if we directly apply the GLRT method introduced in previous section for Fig.~\ref{fig:haystack}(f), we get a confusing false alarm rate map in Fig.~\ref{fig:FAR}(a); it is hard to distinguish Earth from the exozodiacal dust.

\begin{figure}[ht!]
\begin{center}
\begin{tabular}{c}
\includegraphics[width=0.9\textwidth]{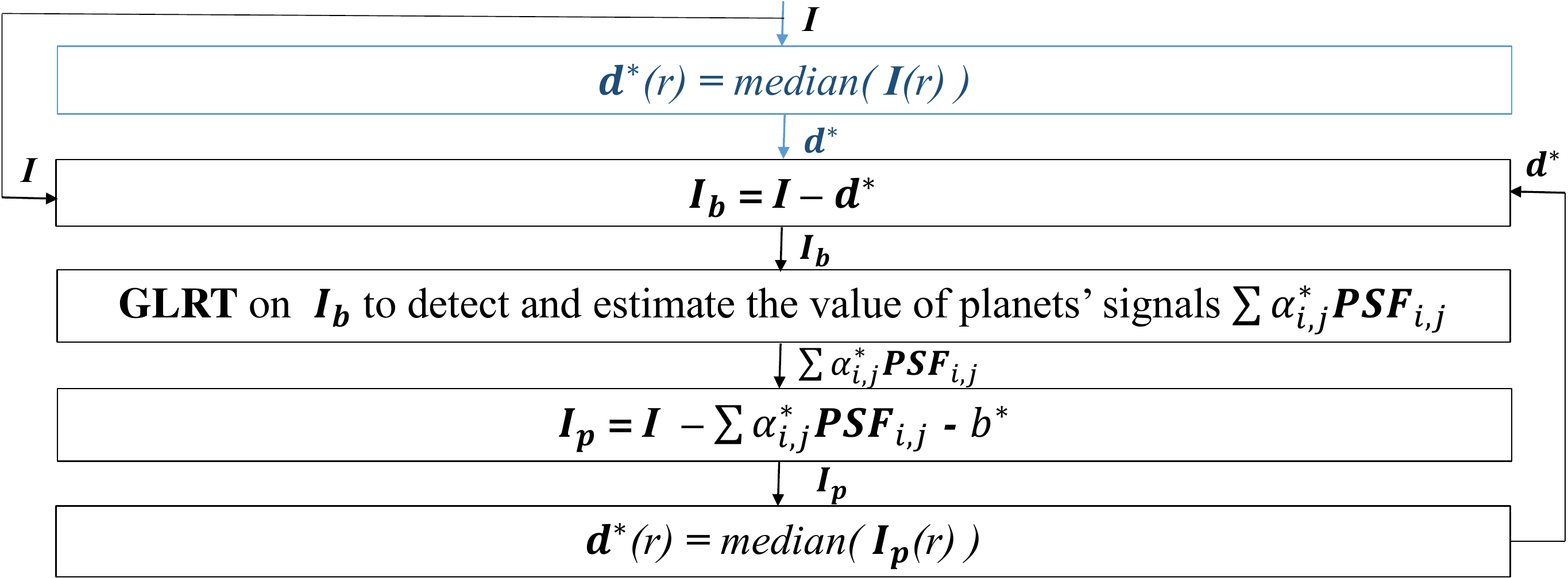}
\end{tabular}
\end{center}
\caption{ The flow chart describes the process of iterative GLRT. The blue box is the initialization step.\label{fig:em}}
\end{figure}

\begin{figure}[ht!]
\centering
\includegraphics[width=0.99 \textwidth,trim= 0  0  0 2cm ,clip]{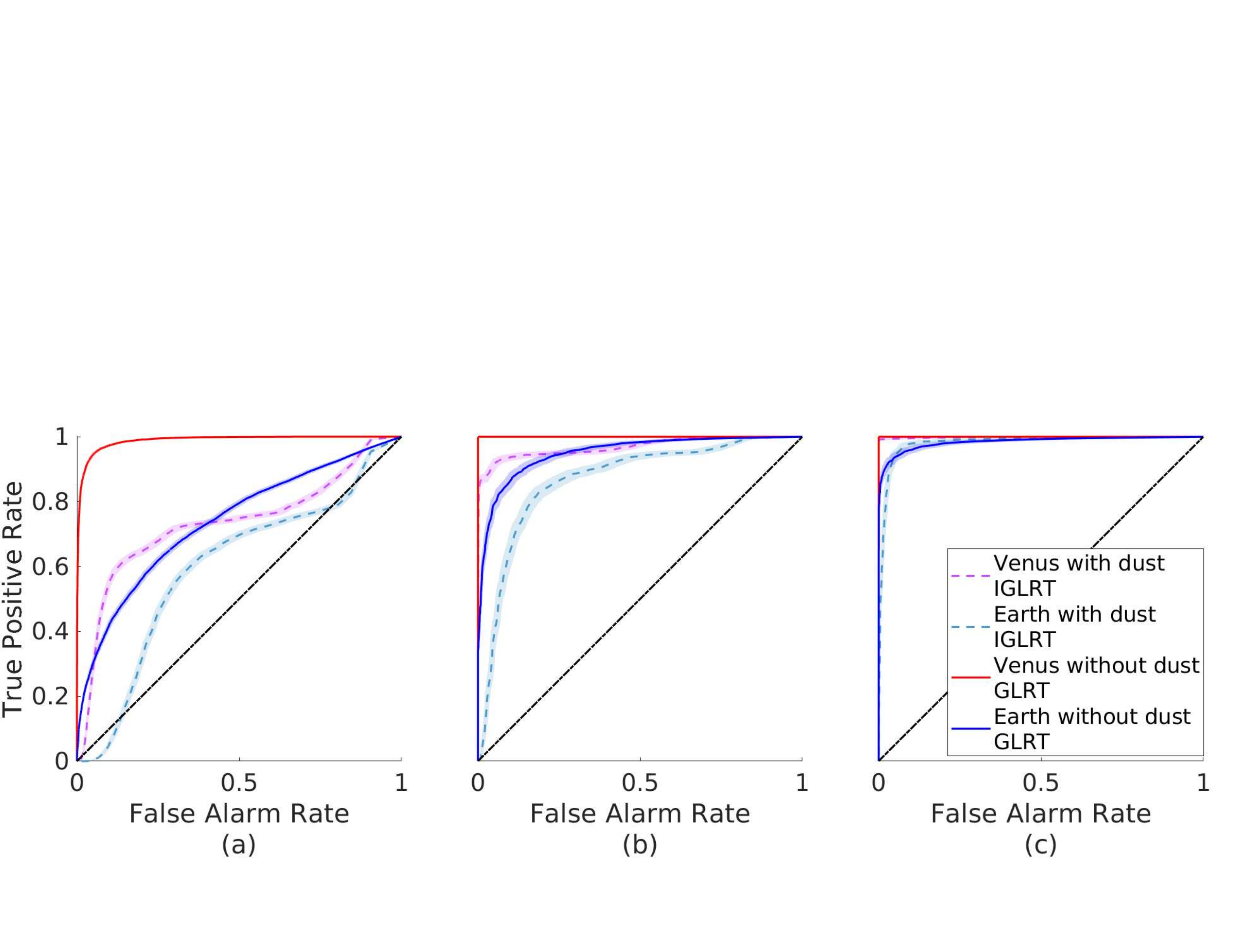}
\caption{  The ROCs for the iterative GLRT.  The ROCs without dust from Fig.~\ref{fig:roc} are added here for reference. The integration time here is 1 second.  (a) ROC for co-added images each from $N_{im}=200$ PC images.  (b) ROC for co-added images each from $N_{im}=700$ PC images. (c) ROC for co-added images each from $N_{im}=2000$ PC images. }
\label{fig:iter_ROC}
\end{figure}

To reduce the number of unknowns, we assume that the dust is nearly axisymmetric, which may be a reasonable approximation for small dust grains \cite{sysmdust}. Thus,
\begin{equation}
\label{model_dust1}
\bm{I}(x,y)= \sum_{i=1}^{N_x}\sum_{j=1}^{N_y}[\alpha_{i,j}  \bm{P}_{i,j}(x,y)] + \bm{ b}(x,y) + \bm{d}(r)+\bm{\omega}(x,y)\,,
\end{equation}
where $r=\sqrt{x^2+y^2}$.
Then Eq.~(\ref{opt}) becomes
\begin{equation}
\label{opt1}
\smash{\displaystyle\min_{\bm{\alpha}, \bm{b},\bm{d} }} ||\bm{I} - \sum_{i =1}^{N_x}\sum_{j =1}^{N_y}[ \alpha_{i,j}   \bm{P}_{i,j}] -\bm{ b}-\bm{d} ||_2\,.
\end{equation}
We can not directly split the whole image into smaller areas and do detection separately like Eq.~(\ref{opt_area}) as the estimation of dust in one area also depends on other areas at the same radii from the center. To tackle this, we split the estimation of signals and exozodiacal dust into two steps. First, we take the median of the values for each radius to estimate the background. We use the median rather than mean to avoid the influence of the existence of planet signals at some radius, shown in Fig.~\ref{fig:FAR}(b). This is equivalent to solving for the optimization problem for each r:
\begin{equation}
\label{opt2}
\bm{d}^{*}(r)=\operatorname*{argmin}_{\bm{d}(r)}||\bm{I}(r) -  \bm{d}(r) ||_1\,.
\end{equation}
 The $*$ denotes the estimate of the corresponding parameter. Then, we subtract the estimated background which contains bright exozodiacal dust, \begin{equation}
\label{opt3}
\bm{I_b}= \bm{I}-\bm{d }^{*}(r)\,.
\end{equation}
An example is shown in Fig.~\ref{fig:FAR}(c). Then, applying GLRT on this image $\bm{I_b}$ produces the  T value map in Fig.~\ref{fig:FAR}(d) and provides an estimation of the planets' positions and intensities, as shown in Fig.~\ref{fig:FAR}(e). The estimated planets are subtracted to get a better estimation of the background,  as shown in Fig.~\ref{fig:FAR}(f) and the process is repeated iteratively. The procedure is summarized in Fig.~\ref{fig:em} and  the complete example is shown in Fig.~\ref{fig:FAR}. In Table~\ref{tab:table}, we summarize the intensity and position estimation error for the example. The ROC curves are shown in Fig.~\ref{fig:iter_ROC}. The performance is undermined a little by the dust, compared to that without dust.

\begin{table}[ht]
\caption{Intensity and position estimation error for  Fig.~\ref{fig:haystack}(f) via iterative GLRT methods} 
\label{tab:table}
\begin{center}       
\begin{tabular}{|l|l|l|} 
\hline
\rule[-1ex]{0pt}{3.5ex} Planet &   Intensity Error    &   Position Error\\
\hline
\rule[-1ex]{0pt}{3.5ex}  Venus &	-5.9\% &   21 milli-arcsec\\
\hline
\rule[-1ex]{0pt}{3.5ex}  Earth&	-38.3\%    &   30 milli-arcsec\\
\hline
\end{tabular}
\end{center}
\end{table}

\section{Conclusion} \label{sec:C}
  A starshade is a promising instrument for the direct imaging of earth-like planets. In this paper, we briefly describe our process for simulating   realistic starshade images and   preliminary study of signal detection in starshade images, which no previous work has looked into. The core detection and estimation part is done by GLRT. We first obtain intensity estimates by QMLE. Then, the likelihood ratio with respect to estimated parameters is calculated. After choosing  a false alarm rate, we can threshold the image and detect the planets. For cases with exozodiacal dust, we split the process into two parts: dust estimation and signal estimation and use GLRT iteratively. Examples using these methods are shown in Sec.~\ref{sec:glrt} and Sec.~\ref{subsec:dust}. The GLRT method successfully and efficiently flags potential planets with a concrete false alarm rate.  It can help distinguish planet signals from  artifacts caused by small starshade shape errors, such as a truncated petal tip. In addition, we provide a guidance to choose the best number of PC images to combine into one co-added image $N_{im}$, utilizing the ROC curves. This will help utilize the observation time efficiently.  
 
 Due to the limitation of Gaussian approximation for the noise distribution in the image, Gaussian GLRT introduces detection  performance improvement but not drastically, compared to the SNR method commonly used in high-contrast imaging. We have worked on an improved version of the GLRT method based on the accurate model for PC images rather than approximation, and thus advances the detection performance\cite{bi_GLRT}; we present the most recent result on this in Ref.~\citenum{Bi_JATIS}. The performance can be further improved if we have prior knowledge about the probability distribution of the planets' intensity in Eq.~(\ref{lr}), which may be available after future exoplanet surveys. In this work, we present the iterative GLRT assuming face-on, uniform exozodiacal dust, but the same concept can be applied to more detailed models of the dust structure.  In this work, spectral information is not discussed. However, the method introduced in this paper can be applicable to different cases. If images at different wavelengths are taken, the product of the likelihood at different wavelengths will be the final likelihood for these images. Then MLE and GLRT can be calculated, and thus detection decision can be made.

\subsection*{Disclosures}
AH is a guest co-editor of this starshade special section.

\subsection* {Acknowledgments}
This work is supported by Caltech-JPL NASA grant NNN12AA01C. The authors thank the anonymous reviewers for their many helpful comments and suggestions.


\bibliography{report}   
\bibliographystyle{spiejour}   


\vspace{2ex}\noindent\textbf{Mengya Hu} is a Ph.D. candidate in Mechanical and Aerospace Engineering, Princeton University. Her research focuses on the image simulation and image processing of space telescope systems with starshades. She graduated from Department of Thermal Science and Energy Engineering, University of Science and Technology of China in 2015 and was awarded the highest honor of the university, the “Guo Mo-Ruo Scholarship”.

\vspace{2ex}\noindent\textbf{He Sun} is a postdoctoral researcher in Computing and Mathematical Sciences (CMS) at the California Institute of Technology. He received his PhD from Princeton University in 2019 and his BS degree from Peking University in 2014. His research focuses on adaptive optics and computational imaging, especially their applications in astrophysical and biomedical sciences, such as exoplanet and black hole imaging.

\vspace{2ex}\noindent\textbf{ N. Jeremy Kasdin} is the Assistant Dean for Engineering at the University of San Francisco and the Eugene Higgins professor of Mechanical and Aerospace Engineering, emeritus, at Princeton University. He received his Ph.D. in 1991 from Stanford University. After being the chief systems engineer for NASA's Gravity Probe B spacecraft, he joined the Princeton faculty in 1999 where he researched high-contrast imaging technology for exoplanet imaging. From 2014 to 2016 he was Vice Dean of the School of Engineering and Applied Science at Princeton. He is currently the Adjutant Scientist for the coronagraph instrument on NASA’s Wide Field Infrared Survey Telescope. 

\vspace{2ex}\noindent\textbf{Anthony Harness} is an Associate Research Scholar in the Mechanical and Aerospace Engineering Department at Princeton University. He received his Ph.D. in Astrophysics in 2016 from the University of Colorado Boulder. He currently leads the experiments at Princeton validating starshade optical technologies. 
\listoffigures
\listoftables

\end{spacing}
\end{document}